\documentclass[12pt,a4paper]{article}
\usepackage[utf8]{inputenc}
\usepackage[T1]{fontenc}
\usepackage{amsmath,amssymb,amsfonts}
\usepackage{physics}
\usepackage{geometry}
\usepackage{enumitem}
\usepackage{hyperref}
\usepackage{braket}

\title{\textbf{Quantum Mechanics:\\[4pt] Problems and Paradoxes}}
\author{L.\,V. Prokhorov\\[4pt]
\small Translated by Alexander S.\ Ushakov and Yury Berdinsky\\[6pt]
\small Saint Petersburg, 2003\\[2pt]
\small SPb.: NIIKH SPbGU, 2003. --- 120 pp.\\[2pt]
\small ISBN 5-7997-0506-8}
\date{}

\begin{document}
\maketitle

\begin{abstract}
This book examines a number of problems of quantum mechanics, most of which
are not usually discussed (What is the origin of probabilities in the
mechanics of the microworld?  What is the nature of Planck's constant $h$?
What is the nature of probability amplitudes?  What is the wave function?
etc.).  A system of axioms for quantum theory is formulated.  A model is
studied according to which a classical oscillator in a thermostat can be
interpreted as a quantum one.  The measurement problem is discussed in
detail.

For advanced undergraduate students, graduate students, and specialists
interested in the foundations of quantum theory.
\end{abstract}

\tableofcontents
\newpage

%======================================================================
\section{Introduction}
\label{sec:intro}
%======================================================================

The last decade has witnessed an unprecedented interest in the foundations of
quantum mechanics.  Reviews have appeared~\cite{Goldstein,Kilin,Mensky2000,Laloe},
discussions are organized in journals~\cite{Haroche,Lipkin} (see also earlier
discussions~\cite{QMdebate}).  Experimental work is carried out with
extraordinary intensity~\cite{Tittel,Weihs,Cornell,Friedman,Nesvizhevsky}.
Possibly this is partly related to the slowdown in the development of
high-energy physics (due to the cancellation of several important experimental
projects).  But primarily the emerging interest should be attributed to the
overall progress of physics.  The period of its extensive development, when
quantum mechanics was applied to ever new problems, has passed.  The time has
come to comprehend the new science, to analyze its foundations with the
accumulated information in mind.

Practical prospects for exploiting the peculiarities of quantum mechanics also
play an important role: quantum computers, the phenomenon of teleportation,
quantum cryptography.  They owe their appearance to the unique capabilities
inherent to quantum mechanics and require a clear understanding of its
foundations.

The successes of quantum theory are enormous; every concrete application
turns out to be successful, and its predictions are confirmed by experiment
with unheard-of precision.

The mathematical apparatus of quantum mechanics is transparent.  Mastering it
usually causes no difficulties, even taking into account subtleties often
ignored by physicists (such as Hermiticity, symmetry, and self-adjointness
of operators).

\medskip
\noindent\textbf{A note on terminology.}\footnote{Traditionally the term
``interpretation'' of quantum mechanics has been established, i.e.\ the
explication of the concepts employed by it.  What should really be discussed
is ``understanding'', i.e.\ its \emph{modeling}.  The theory of phlogiston
also needed an ``interpretation''; it explained the incomprehensible
through the unknown.}

Questions are generated both by the physical consequences flowing from the
mathematical apparatus of quantum theory --- wave--particle duality,
uncertainty relations, etc., which are fully confirmed by experiment --- and
by the absence of a clear understanding of the physical nature of the wave
function.  All this engenders a feeling of dissatisfaction that remains after
studying it.  But while previously expressing such dissatisfaction was a
prejudicial and, perhaps, even a somewhat improper matter (it was considered
a sign that one simply could not understand ``the new physics''), now people
speak about it openly.  At the present time it has become clear that a
number of concepts with which orthodox (``Copenhagen'' or ``G\"ottingen'')
quantum mechanics operated (point particles, uncontrollable interaction of
the object with the apparatus, ``non-causal change'' of the wave
function~\cite{Messiah}, etc.) need refinement; that the fundamental
questions generated by quantum mechanics (the nature of probability
amplitudes, the nature of Planck's constant $h$, the origin of probabilities
in mechanics, etc.) are legitimate and require answers.  By the beginning of
the 21st century an enormous experimental record has been accumulated, a
gigantic step forward has been made in theory --- from black holes to the
model of grand unification.  Intensive research is being conducted on physics
at Planck distances $\sim l_P \approx 1.6\times 10^{-33}\;\text{cm}$
(superstring theory).  A most serious challenge to quantum theory is posed
by the theory of gravitation.  Besides the non-renormalizability of the
gravitational Lagrangian, there exists the even more important problem of
black holes --- in them information is lost~\cite{Hawking}, which is
impossible in a theory with a self-adjoint Hamiltonian.  All this testifies
to the necessity of a critical analysis of the foundations of quantum
theory.\footnote{And once more about terminology.  By \emph{quantum
mechanics} (QM) we shall mean the quantum theory of mechanical systems with
a finite number of degrees of freedom; by \emph{quantum field theory}
(QFT) --- the theory of systems with an infinite number of degrees of
freedom.  A theory based on the concept of probability amplitude will be
called \emph{quantum theory}.}\textsuperscript{,}\footnote{Serious attempts
to understand the meaning of Planck's constant $h$ were undertaken,
apparently, only at the very earliest stages of quantum theory, before the
creation of quantum mechanics~\cite{Jammer}.}

\subsection*{Questions generated by quantum theory}
\subsubsection*{Questions that existed from the very beginning}

Problematic questions generated by quantum mechanics can be divided into
three groups: general, particular, and questions related to the measurement
problem.  General questions pertain to any quantum theory and are usually
not discussed (and not even posed).

\medskip
\noindent\textbf{Group~I.}
\begin{enumerate}
\item What is the origin of probabilities in mechanics?
\item What is the nature of probability amplitudes?
\item What is the nature of Planck's constant $h$?
\end{enumerate}

The answers to these touch not so much on one or another theory, as on the
very foundations of the universe.  Nevertheless, they all lie within the
competence of physics.

Particular questions are more concrete.  They pertain to quantum mechanics
and primarily concern unusual properties of objects in the microworld.

\medskip
\noindent\textbf{Group~II.}
\begin{enumerate}
\item How are classical probability theory and the probabilistic
mathematical apparatus of quantum theory related to each other?
\item What is a particle?
\item What physical reality corresponds to the wave function?
\item What is the nature of wave--particle duality?
\item What is the nature of the uncertainty relations?
\item Is the quantum-mechanical description complete?
\end{enumerate}

The first question is not a question of physics but of mathematics.  The
answers to the remaining questions cannot be given within the framework of
QM, since, for example, the concept of the wave function and its properties
are postulated.  But only knowing them can one hope to explain the
point-like character of particles (the possibility of detecting them in an
arbitrarily small region of space) and the integrity of quanta
(single-particle excitations of fields).  One should expect that answers to
most questions will appear after elucidating the nature of the wave
function, i.e.\ elucidating the physics that stands behind it.

A specific place in quantum theory is occupied by questions related to
observation and measurement, i.e.\ to the experimental study of properties
of micro-objects and processes involving them.  This group of questions is
conditioned, on the one hand, by the radical difference of quantum mechanics
from classical mechanics, and on the other hand, by the classical character
of the apparatus used by the experimenter.\footnote{According
to~\cite{EPR}: ``\ldots every element of physical reality must have a
counterpart in physical theory.''}

\medskip
\noindent\textbf{Group~III.}
\begin{enumerate}
\item Do macroscopic bodies (apparatus) obey the laws of quantum mechanics?
\item What should be considered an act of measurement?
\item Can the act of measurement be described?
\item What is the role of the observer?
\end{enumerate}

In classical physics such questions do not arise.  In quantum physics there
are no universally accepted answers to them to this day.  Hence the
appearance of such concepts as ``quantum jumps'', ``non-causal change'' of
the wave function during measurement, ``uncontrollable interaction'' of the
object with the apparatus.

The presence of so many important questions testifies to the incompleteness
of QM as a theory of the microworld.  This is precisely the reason for the
appearance of various paradoxes and numerous interpretations of quantum
mechanics.

\subsubsection*{Questions that appeared with time}

As quantum physics developed, entirely unexpected problems were discovered
that were impossible to foresee at the initial stage of its formation.  All
of them are in one way or another connected with the peculiarities of the
structure of space-time.

\medskip
\noindent\textbf{Group~IV.}
\begin{enumerate}
\item Ultraviolet divergences in the theory of quantized fields.
\item Non-renormalizability of gravitational interaction.
\item Loss of information in black holes.
\item Sign-indefiniteness of the probability density for bosons in
relativistic theory: the ``density''
\begin{equation*}
\rho(x)=i\varphi^+\!\overset{\leftrightarrow}{\partial_0}\varphi^+
= i(\varphi^+\partial_0\varphi^+ -\partial_0\varphi^+\cdot\varphi^+),
\end{equation*}
where $\varphi^+$ is a solution of the Klein--Fock--Gordon (KFG) equation
with positive energy, is not positive definite (the fact itself was known
already in the period of the creation of quantum mechanics --- it was
precisely for this reason that the KFG equation was initially rejected; the
problem reappeared after quantization of the scalar field).
\end{enumerate}

The question of the existence of a self-consistent quantum mechanics of
bosons in Minkowski space is usually answered in the
negative~\cite{BLP}.  This evidently testifies to the fact that the
problems of quantum theory are closely tied to principal problems of the
structure of space at small distances.  The same is testified to by the
difficulties of quantizing fields with local interaction, especially
quantization of the gravitational field, which clearly points to the
insufficiency of existing conceptions of space-time in the small.  The
problem of loss of information in black holes makes the situation absolutely
unacceptable.

\subsubsection*{Brief commentary}

The questions of Group~I concern fundamental properties of quantum theory.
Answers to them can be obtained only after constructing a model that
operates with probability amplitudes and explains the reason for the
appearance of a universal constant having the dimension of action.  The
existence of such a constant is all the more remarkable because action is
not a conserved quantity.  The appearance of $h$ testifies to the existence
of deep connections between space and material objects at ultra-small
(presumably Planckian) distances.  The fact that $h$ also has the dimension
of angular momentum, which is conserved, is immaterial, since quantum theory
can be formulated in one-dimensional space as well.  But $h$ has the
dimension of \emph{area of phase space} (PS).\footnote{The term ``volume''
will also be applied to two-dimensional PS.}
It is precisely this fact that turns out to be decisive in modeling QM.  In
Section~4 a model is proposed in which Planck's constant has a clear meaning
--- it is the volume of the phase spaces of the elementary oscillators
forming the field.  Thus, one should expect that exhaustive answers will be
obtained only upon transition to a more general theory, i.e.\ upon
reaching the boundary beyond which quantum theory and the concept of
continuous space lose their force.

The first question of the second group stands apart.  The answer to it
should be given by mathematics.  Probability theory, which took shape as a
mathematical discipline also in the 20th century, is based on the axioms of
Kolmogorov~\cite{Kolmogorov}.  Being a deeply developed and widely applied
branch of mathematics, it nevertheless says nothing about probabilistic
schemes resting on the concept of probability amplitude, nor about
processes described by complex amplitudes.  This is surprising, since QM is
an intrinsically probabilistic theory, similar, say, to statistical
physics.  Even more surprising is the fact that although the mathematical
apparatus of QM (the theory of Hilbert spaces) was known even before its
creation, it was never developed as a branch of probability theory operating
with state vectors of dynamical systems.  However, in modern courses on
probability theory, complex random variables are considered and averages of
their products, equivalent to scalar products in a Hilbert space, are
defined~\cite{Loeve}.  Moreover, it turns out that there exists a theorem
admitting the existence of probabilistic schemes with probability
amplitudes.  Classical probability theory operates with non-negative
quantities --- probability distributions.  Now, it is asserted~\cite{Lukacs}
that a non-negative function $p(x)$ is a probability density if and only if
it is the square of the modulus of some complex function $\psi(x)$.  This
partly legitimizes QM as a probabilistic theory, but only partly, since
nothing is said about the rules for handling the functions~$\psi$.

The remaining questions of the second group explicitly or implicitly figure
in the physics literature, and the last of them is widely discussed to this
day (the discussion was initiated by the paper~\cite{EPR}).  The
mathematical apparatus of QM contains no indication of the physical nature
of the wave function or its interrelation with space.  The main difficulty
here is the absence of models of the wave function and of particles.  In
fact, the correct answer to question~II\,2 was known long ago (see, e.g.,
\cite{Blohintsev,Frenkel}): a particle is a quantum (single-particle
excitation) of the corresponding field.  From here a natural conclusion
follows: the wave function is a function describing an excitation of a
field.  The decisive role, as could be expected, is played by the field.
It determines all properties of its excitations, and therefore the answers
to the remaining questions follow almost automatically (see Section~3).

Concepts and ideas formed in the process of creating QM and as a result of
its practical application also become clarified and gain precision when one
turns to QFT\@.  For example, it is clear that if a particle is a quantum of
a field, then it cannot be a point (for more detail, see Section~3).  The
main reason for the difficulties was the impossibility of modeling QM within
the framework of classical mechanics and classical, i.e., Euclidean space
(relativistic theory changes nothing).  It is necessary to model not only
particles but also space, i.e., a radical departure beyond the framework of
classical physics is required (Sections~4,~5).  In this case one succeeds in
introducing $h$ into the theory.  If the speed of light $c$ establishes a
connection between the measures of space and time ($T=L/c$), then Planck's
constant $h$ points to a connection between energy (i.e., a dynamical
characteristic of a particle) and time ($T=h/E$), or momentum and space
($L=h/p$).  From here one can conclude that in QM the connection of dynamics
with space should play an important role (in classical physics it manifests
itself as the property of inertia of bodies).

A special role in QM is played by the measurement problem (questions of the
third group).  An enormous literature is devoted to it.  The difficulty is
that in the act of measurement two heterogeneous objects participate --- a
microparticle, subject to the laws of QM, and a classical apparatus.  Which
mechanics should be used to describe the process of measurement?  The
problem was aggravated by the abstract character of the wave function, the
absence in the apparatus of QM of a mechanism for the transition from
probability amplitudes to probabilities, which led to fantastic suggestions
about the special role of the observer (``consciousness'') in quantum
physics.  The axiomatic formulation of QM (Section~2) and the elucidation
of the peculiarities of the quantum description of macroscopic bodies
(superselection rules, Section~6) allow one to overcome the main
difficulties connected with the procedure of observation in QM.

The questions of the fourth group point to the necessity of a radical
change in our conceptions of space-time at small (Planckian) distances.
Here one should expect the appearance of discreteness (Section~5).  One
can suppose that the decisive role will be played by superstring theory.
Apparently, superstrings are the form of manifestation of matter at Planckian
distances.  If at distances $10^{-1}$--$10^{-16}$~cm the characteristic
structures are molecules, atoms, nuclei, elementary particles, then at
distances $\sim 10^{-33}$~cm such structures will be superstrings.

\medskip
\noindent\textbf{Contents of the book.}

In Section~2 the question of the place that theories based on the concept of
probability amplitude occupy within probability theory is considered.  Rules
for handling probability amplitudes (axioms) are formulated.  Their
consistency and independence are demonstrated.  The existence of such axioms
promotes quantum mechanics to the status of one of the branches of
probability theory.

In Section~3 the consequences of the fact that all known particles are
excitations of one or another field are discussed.  Essentially, this
circumstance allows one to answer all ``particular'' questions of Group~II.
Indeed, in this case the wave function should be identified with a function
describing a single-particle excitation of a field, and the appearance of the
uncertainty relations becomes obvious.  The impossibility of making the
description of particles ``more complete'', i.e., of introducing additional
characteristics of particles beyond those allowed by quantum mechanics, also
becomes obvious.

Section~4 is devoted to questions of the nature of Planck's constant $h$
and its role in quantum theory.  It is shown that a classical theory with a
compact phase space (a sphere) whose area equals $h$ is equivalent to the
system ``harmonic oscillator in a thermostat'', which in turn can be
interpreted as a quantum oscillator if the Gibbs measure is identified with
the volume measure of phase space.  Fock space arises naturally in this
construction.

In Section~5 a chain of classical harmonic oscillators in a thermostat is
considered.  This system models both a one-dimensional space and a
one-dimensional quantum field theory.

Problems of measurement, decoherence, and the quantum description of
macroscopic bodies are discussed in Section~6.  It is shown that for
macroscopically distinct states of macroscopic bodies there exist
superselection rules.  This resolves the main difficulty of the theory of
measurement --- the question of the macroscopic definiteness of apparatus.

In Section~7 some experimental consequences of the assertions of Section~3
are discussed: ``splitting'' of quanta and interference of two photons.

In Section~8 a critical review of the most popular ``interpretations'' of
quantum mechanics is given; frequently encountered questions about causality,
the role of language, and the role of philosophy in understanding quantum
theory are touched upon.

In the concluding Section~9, the questions formulated in the Introduction
are re-analyzed.

In the Appendix (Section~10), information about superselection rules is
provided, a brief exposition of the most important paradoxes is given.  In
Section~10.3 the simplest conceivable model of a quantum theory is
considered.  It is shown that there exist classical theories with a universal
constant of the dimension of action, and that there exist quantum theories
without Planck's constant $h$.  In Section~10.4 some formulas of Section~4
are proved.

%======================================================================
\section{Quantum Mechanics and Probability Theory}
\label{sec:QMprob}
%======================================================================

The questions listed in Section~1 have varying degrees of complexity.  The
questions of the first group are the most complex and profound.  To answer
them, a new concept of physical space is needed.  The questions of the
second group are much simpler and do not require going beyond existing
theories.  We begin with them.

\subsection{Probability amplitudes and classical theory}
\label{sec:2.1}

Let us find out whether in standard probability theory a distribution (or
density) of probability can be specified by a bilinear function of dynamical
variables.  Let the state of a material point be characterized by a
probability density $W(q,p)$, where $q$ and $p$ fix a point in the
two-dimensional phase space ($W$ can also be a probability distribution).
If the densities $w_1(q)$ and $w_2(p)$ on the $q$ and $p$ axes are
independent, then $W(q,p)=w_1(q)\,w_2(p)$.

This circumstance can be written in various forms.  For example, let
$\mathbf{w}(w_1,w_2)$ be a two-dimensional vector with components $w_1$,
$w_2$.  Introducing the metric tensor $g=\sigma_1$ ($\sigma_{1,2,3}$ are
Pauli matrices), we have $W=\mathbf{w}\,g\,\mathbf{w}/2=(\mathbf{w},
\mathbf{w})/2$.  Note the \emph{bilinear} dependence of $W$ on the
two-dimensional vector~$\mathbf{w}$: such dependencies, characteristic of
quantum mechanics, are by no means foreign to classical probability theory.

Other forms of writing are also possible.  Alongside $\mathbf{w}$ we
introduce the conjugate vector $\bar{\mathbf{w}}=\mathbf{w}\sigma_3$.  Then
$W=\bar{\mathbf{w}}\wedge\mathbf{w}/2$, where $\wedge$ symbolizes the
exterior product of vectors, and the density is the measure of the bivector
$|\mathbf{W}|$ (the area defined by the bivector).  There is also yet another
form of writing, more suitable for our purposes than the others.

Hamilton's equations of motion
$\dot{q}=\partial H/\partial p=\{q,H\}$,
$\dot{p}=-\partial H/\partial q=\{p,H\}$
($H$ is the Hamiltonian, $\{\,,\}$ the Poisson brackets) can be written with
the help of complex numbers
$z=(q+ip)/\sqrt{2}$, $\bar{z}=(q-ip)/\sqrt{2}$, $\{z,\bar{z}\}=-i$,
$\dot{z}=-i\,\partial H(z,\bar{z})/\partial\bar{z}$.

If from the functions $w_{1,2}$ one constructs a complex number
\begin{equation*}
\psi = w_1 + iw_2,
\end{equation*}
then $W=|\psi|^2/2=w_1 w_2$ (the components of the 2-vector are taken to be
$w_1$, $iw_2$).  But then one must restrict oneself to the positive quadrant
of the complex $\psi$-plane, since the physical region is defined by the
inequalities $w_{1,2}\ge 0$.  One could use the entire plane $w_1$, $w_2$
only if it were possible to unambiguously define $\psi$ outside the physical
region.  In that case, in intermediate calculations $\psi$ could be treated
as arbitrary, setting $w_{1,2}\ge 0$ only in the final formulas.  For
example, $\psi$ may be an analytic function of~$z$, i.e.,
\begin{equation*}
\psi(z)\,:\qquad \frac{\partial\psi}{\partial\bar{z}}=0.
\end{equation*}

The last equality is a functional equation.  Its solution is given by a
linear function $\psi(z)=bz+c$, $\operatorname{Im} b=0$.  In what follows,
without loss of generality we set $b=1$, $c=0$.  If the Hamiltonian is
invariant under rotations in the complex plane, i.e., if
$H(z_1,\bar{z}_1)=H(z,\bar{z})$, where $z_1=e^{i\alpha}z$, then the
``coordinate axis'' can be chosen arbitrarily.

Thus, a description with the help of complex numbers, extended to the entire
plane, is suitable only for distributions of a very special kind.  The
``complex probability'' $\psi(z)$ is proportional to the canonical variable
$z$, and therefore from the normalizability of the probability it follows
that motion in phase space must be \emph{finite} (as, for example, motion on
the Riemann sphere or on a torus).  There is a duality: the variable $z$
plays a double role --- it is both a dynamical variable and a ``complex
probability'', i.e.,
\begin{equation*}
\text{canonical variable}\;\leftrightarrow\; z
\;\leftrightarrow\;\text{probability amplitude.}
\end{equation*}

The transition to complex canonical variables brings the mathematical
apparatus of classical theory closer to the quantum one.  The construction of
specific models of this type, as well as a discussion of the reasons for the
probabilistic description of a particle, is given in Sections~4,~5.

If the phase space is compact (or if for some other reasons motion in it is
finite), then the meaning of Planck's constant $h$ becomes clear --- it
connects the probability $P$ with the area of phase space ($P\sim|qp|/h$).
Furthermore, if $z$ is a dimensionless quantity, then the existence of
another world constant ($l$) of the dimension of length is necessary; then
$z=(q/l+ilp/\hbar)/\sqrt{2}$.  The constant $l$ is naturally identified
with the Planck length: $l\sim l_P$.  Let us also note that the formalism
can be made entirely real by replacing $i\to -\sigma_2$.

In order to determine the probability in the general case with a ``complex''
coordinate axis, i.e., when $q$ and $p$ are complex numbers, we introduce
the vector $\mathbf{z}(z_q,z_p)=z_q\mathbf{e}_q+z_p\mathbf{e}_p$, where
$z_q$ and $z_p$ are complex numbers and $\mathbf{e}_{q,p}$ are basis vectors
in phase space.  Then
\begin{equation*}
W = i\bar{\mathbf{z}}\wedge\mathbf{z}
  = i(z_q^* z_p - z_p^* z_q)\,\mathbf{e},
\end{equation*}
where $\mathbf{e}=\mathbf{e}_q\wedge\mathbf{e}_p$ is the unit bivector.
The probability is given by the \emph{coefficient} at~$\mathbf{e}$.

\medskip
\noindent\textbf{Arbitrary dimensions.}
The results obtained are easily generalized to the case of a space of
arbitrary dimension~$n$.  Let
\begin{equation*}
W(x_1,\ldots,x_n) = w_1(x_1)\cdots w_n(x_n);
\end{equation*}
form the vectors
\begin{equation*}
\mathbf{w}_i = \sum_{j=1}^{2} \alpha_{ij}\, w_j\, \mathbf{e}_j,
\end{equation*}
where $\alpha_{ij}$ are elements of some non-degenerate matrix,
$\det\alpha\ne 0$.  Then
\begin{equation*}
W = \frac{1}{\det\alpha}\,
    \mathbf{w}_1\wedge\cdots\wedge\mathbf{w}_n
  = w_1(x_1)\cdots w_n(x_n)\,\mathbf{e}_1\wedge\cdots\wedge\mathbf{e}_n
  = W\,\mathbf{e}.
\end{equation*}

\medskip
\noindent\textbf{Arbitrary densities.}
The requirement of factorizability of $W$ is not obligatory.  Suppose that
\begin{equation*}
W(x,y) = \sum_k w_1^{(k)}(x)\,w_2^{(k)}(y).
\end{equation*}
One can form the vector
\begin{equation*}
\mathbf{w} = \sum_{i=1,2}\sum_k w_i^{(k)}\,\mathbf{e}_i^{(k)},
\end{equation*}
where
\begin{equation*}
\mathbf{e}_i^{(k)} = \mathbf{e}_i\,\xi^{(k)},\qquad
\mathbf{e}_i^{(k)}\wedge\mathbf{e}_j^{(k')}
= \mathbf{e}_i\wedge\mathbf{e}_j\cdot\xi^{(k)}\cdot\xi^{(k')},
\end{equation*}
\begin{equation*}
\mathbf{e}_i\wedge\mathbf{e}_j = \epsilon_{ij}\,\mathbf{e},\qquad
\xi^{(k)}\cdot\xi^{(k')} = \delta_{kk'},\qquad
\epsilon_{12}=1,
\end{equation*}
and the conjugate vector
$\bar{\mathbf{w}}=\sum_{i=1,2}\sum_k(-1)^{i/2}w_i^{(k)}\,
\bar{\mathbf{e}}_i^{(k)}$.

We have:
\begin{equation*}
W = \tfrac{1}{2}\,\bar{\mathbf{w}}\wedge\mathbf{w}
  = \sum_k w_1^{(k)}\,w_2^{(k)}\,\mathbf{e} = W\,\mathbf{e},
\end{equation*}
i.e., in this case too the probabilities are given by bilinear expressions
in auxiliary quantities.  The generalization of the obtained formulas to the
case of a space of arbitrary dimension is trivial.

\medskip
\noindent\textbf{Complex variables.}
If the variables $q$, $p$ are initially complex (i.e., if $q$ describes a
system with two degrees of freedom), then
$q^{(\pm)}=(q_1\pm iq_2)/\sqrt{2}$, $p^{(\pm)}=(p_1\mp ip_2)/\sqrt{2}$,
$\{q^{(+)},p^{(+)}\}=1$,
$\mathbf{z}^{(+)}=q^{(+)}\mathbf{e}_q+p^{(+)}\mathbf{e}_p$, and
\begin{equation*}
\mathbf{w}^{(+)} = i\,\bar{\mathbf{z}}^{(+)}\wedge\mathbf{z}^{(+)}
= i\bigl[q^{(+)*}\!\wedge p^{(+)}\,\mathbf{e}_q\wedge\mathbf{e}_p
  + p^{(+)*}\!\wedge q^{(+)}\,\mathbf{e}_p\wedge\mathbf{e}_q\bigr].
\end{equation*}
Here the exterior product $q^{(+)*}\wedge p^{(+)}$ is the product of
vectors $q^{(+)*}$ and $p^{(+)}$ with components $(q_1/\sqrt{2},
-iq_2/\sqrt{2})$ and $(p_1/\sqrt{2},-ip_2/\sqrt{2})$, i.e.,
$q^{(+)*}\wedge p^{(+)}=-i(q_1 p_2-q_2 p_1)/2$.

\medskip
\noindent\textbf{Fermions.}
Let $\psi=q\theta/\sqrt{2}$ and $\psi^+=q^*\bar{\theta}/\sqrt{2}$, where $q$
is a complex number, $\theta=\theta_1+i\theta_2$,
$\bar{\theta}=\theta_1-i\theta_2$, and $\theta_{1,2}$ are real generators of
a Grassmann algebra, $[\theta_1,\theta_2]_+=0$.  In this case the Lagrangian
is linear in velocities ($\dot{\mathcal{L}}\sim\psi^+\dot{\psi}$), so that
$p_\psi=\psi^+$.  As $\mathbf{z}$ take
$\mathbf{z}=\psi\,\mathbf{e}_q+\psi^+\mathbf{e}_p$.  Define:
\begin{equation*}
W = \bar{\mathbf{z}}\wedge\mathbf{z}
  = \psi^*\psi^+\,\mathbf{e}_q\wedge\mathbf{e}_p
  + \psi^{+*}\psi\,\mathbf{e}_p\wedge\mathbf{e}_q
  = qq^*\,\bar{\theta}\theta\;\mathbf{e},
\end{equation*}
where $\bar{\theta}\theta=\theta\bar{\theta}=2i\theta_2\theta_1$; then the
probability is the coefficient at $\bar{\theta}\theta\,\mathbf{e}$.

Thus, classical theory admits constructions in which probabilities are given
by bilinear expressions of complex functions.  Among the questions of
Group~II the first turns out to be the simplest.  For the answer to it one
does not need to invent new models --- it is a question not of physics but
of mathematics.  What then does mathematics say on this matter?

\subsection{Theorem on the characteristic function}
\label{sec:2.2}

Does there exist a probabilistic theory operating with probability
amplitudes, and if so, what place does it occupy in probability theory?
Although the mathematical apparatus of QM was developed in detail even
before its discovery (theory of Hilbert spaces, theory of linear partial
differential equations), it was never connected with classical probability
theory, and QM was never regarded as one of its branches.  Undoubtedly, this
is one of the reasons that hamper the understanding of QM\@.  Without
absolute mathematical clarity it is difficult to make sense of sufficiently
complicated physics.  Fortunately, the first important step in this direction
has been made.

It turns out that standard probability theory not only admits the existence
of mathematical objects with the properties of probability amplitudes, but
also proves the local equivalence of the probability density $p(x)$ and
$|\psi(x)|^2$.  The following theorem holds~\cite{Lukacs}:

\medskip
\noindent\textbf{THEOREM.}
\emph{A complex-valued function $f(t)$ of a real variable $t$ is the
characteristic function of an absolutely continuous distribution if and only
if it admits the representation}
\begin{equation}\label{eq:2.1}
f(t) = \int_{-\infty}^{\infty} g(t+\xi)\,\overline{g(\xi)}\,d\xi,
\end{equation}
\emph{where $g(\xi)$ is a complex function of a real variable $\xi$ such
that}
\begin{equation*}
\int_{-\infty}^{\infty}|g(\xi)|^2\,d\xi < \infty.
\end{equation*}

\medskip
Recall that a distribution is absolutely continuous if it is given by a
probability density $p(x)\ge 0$.  The characteristic function $f(t)$ is the
Fourier transform of the function $p(x)$:
\begin{equation*}
f(t) = \int_{-\infty}^{\infty} dx\, e^{itx}\,p(x).
\end{equation*}

Passing to a new function $\psi(x)$:
\begin{equation*}
g(\xi) = (2\pi)^{-1/2}\int_{-\infty}^{\infty} dx\,\psi(x)\,e^{i\xi x},
\end{equation*}
we rewrite~\eqref{eq:2.1} in the form
\begin{equation}\label{eq:2.2}
p(x) = |\psi(x)|^2.
\end{equation}
The equality~\eqref{eq:2.2} establishes a connection between the
mathematical apparatus of quantum mechanics and the apparatus of probability
theory (in connection with QM, this theorem was also discussed
in~\cite{Cohen}).  From~\eqref{eq:2.2} it follows that for any absolutely
continuous distribution on the real axis the probability density can always
be represented as the square of the modulus of some complex function from
$L_2$, i.e., a function belonging to a Hilbert space.  And conversely,
$|\psi(x)|^2$, where $\psi\in L_2$, always defines some absolutely
continuous probability distribution.  But it is precisely such functions
that are used in quantum mechanics.  From here it follows that the
quantum-mechanical probabilistic description fits entirely within the
standard scheme of probability theory defined by the axioms of
A.\,N.~Kolmogorov~\cite{Kolmogorov}.  It does not narrow the standard
theory.

However, the foregoing does not mean that with the transition to complex
functions $\psi(x)$ (probability amplitudes) nothing new appears.  If one
considers that the functions $\psi(x)$ correspond to some ``physical
reality'' and takes them as the basis of a physical theory, then it is
necessary to specify rules for handling probability amplitudes.  Obviously,
they must be such that the rules for handling probability distributions
following from them do not contradict the axioms of
Kolmogorov~\cite{Kolmogorov}.  The axioms of a theory based on the concept
of probability amplitude are formulated in~\cite{Prokhorov1975}.

\subsection{Axioms of quantum theory}
\label{sec:2.3}

Let us formulate the rules for handling probability amplitudes.  For
didactic reasons, let us begin with the axioms of classical probability
theory.

\medskip
\noindent\textbf{Axioms of classical probability theory.}

\medskip
\noindent\textsc{Definitions.}

\noindent\textbf{O1.}\ There is a certain set $\Omega$ of elements $\omega$,
called the set of \emph{elementary events}.

\noindent\textbf{O2.}\ There is a Borel field $\mathcal{A}$ of subsets of
the set $\Omega$.  The elements of $\mathcal{A}$ are called
\emph{random events}.

\medskip
\noindent\textsc{Properties of the field $\mathcal{A}$.}

\noindent (i) $\Omega\in\mathcal{A}$.

\noindent (ii) If $A\subset\Omega$, $B\subset\Omega$ and
$A\in\mathcal{A}$, $B\in\mathcal{A}$, then $A+B\in\mathcal{A}$,
$AB\in\mathcal{A}$, $\bar{A}\in\mathcal{A}$, $\bar{B}\in\mathcal{A}$.

Here $A+B$ is the union of sets $A$ and $B$, $AB$ is their intersection,
$\bar{A}$ is the complement to the set $A$.  Obviously,
$\bar{\Omega}=\varnothing\in\mathcal{A}$, where $\varnothing$ is the
empty set.

\noindent (iii) If $A_i\subset\Omega$ ($i=1,2,\ldots$) and
$A_i\in\mathcal{A}$, then $A_1+A_2+\cdots\in\mathcal{A}$ and
$A_1 A_2\cdots\in\mathcal{A}$.

The set $\mathcal{A}$ is called the \emph{field of events}.

\medskip
\noindent\textsc{Axioms.}

\noindent\textbf{C1.}\ To each random event $A$ from the field $\mathcal{A}$
there is assigned a certain non-negative number $P(A)$ (probability of the
event).

\noindent\textbf{C2.}\ If events $A_1,A_2,\ldots,A_n$ are pairwise
mutually exclusive (i.e., $A_i A_k=\varnothing$), then
$P(A_1+A_2+\cdots+A_n)=P(A_1)+P(A_2)+\cdots+P(A_n)$, even if
$n\to\infty$.

\noindent\textbf{C3.}\ $P(\Omega)=1$.

The collection $\{\Omega,\mathcal{A},P\}$ is called a \emph{probability
space}.

\medskip
Now let us formulate the quantum axioms.

\medskip
\noindent\textbf{Axioms of quantum theory}~\cite{Prokhorov1975}.

\medskip
\noindent\textsc{Definitions.}

\noindent\textbf{O1.}\ There is a triple of sets $E$, $\bar{E}$, $\Omega$
with elements denoted respectively $e_i$, $\bar{e}_j$, called
\emph{states}, and $\omega_j^i$, called \emph{elementary events}.

\noindent\textbf{O2.}\ There is a Borel field $\mathcal{A}$ of subsets of
the set $\Omega$.  The elements of the sets $E$, $\bar{E}$, $\mathcal{A}$
form the set $\mathcal{U}$.

\medskip
\noindent\textsc{Axioms.}

\noindent\textbf{Q1.}\ The elements of the sets $E$, $\bar{E}$, $\Omega$
are in mutual one-to-one correspondence, whereby the element $e_i$ is
associated with $\bar{e}_i$, and to the pair of elements $\bar{e}_i$, $e_j$
is associated the element $\omega_j^i\equiv\bar{e}_i e_j\equiv
\bar{e}_j e_i$.

\noindent\textbf{Q2.}\ To each element $u$ of the set $\mathcal{U}$ there
is assigned a complex number $\psi(u)$:
$e_i\to\psi(e_i)$, $\bar{e}_j\to\psi(\bar{e}_j)$,
$\omega_j^i\to\psi(\omega_j^i)=\psi(\bar{e}_i)\psi(e_j)$.

\noindent\textbf{Q3.}\ If $A\in\mathcal{A}$, $B\in\mathcal{A}$ and
$AB=\varnothing$, then $\psi(A+B)=\psi(A)+\psi(B)$.

\noindent\textbf{Q4.}\ $\psi(\Omega)=1$.

\medskip
The axioms are formulated.  They are \emph{independent} (no axiom is a
consequence of the others) and \emph{consistent}.  The proof of the latter
reduces to indicating an example in which all axioms are fulfilled.
\emph{Example}: the sets $E$, $\bar{E}$, $\Omega$ consist of one element
each, respectively $e$, $\bar{e}$ and $\omega$; $\mathcal{A}$ consists of
$\omega$ and $\varnothing$; $\psi(e)=e^{i\alpha}$,
$\psi(\bar{e})=e^{-i\alpha}$, $\psi(\omega)=1$, $\psi(\varnothing)=0$;
$\operatorname{Im}\alpha=0$.

As in classical theory, this system of axioms is \emph{incomplete} (for
example, the functions $\psi(e_i)$, $\psi(\bar{e}_j)$ are not fixed).  But
it is incomplete for another reason as well.

This system of axioms defines \emph{some} quantum theory, since complex
probability amplitudes $\psi$ have been introduced.  But it does not
indicate the rules for handling them, so let us add two more axioms.

\noindent\textbf{Q5.}\ The sets of states $\{e_i\}$, $\{\bar{e}_j\}$ form
complex vector spaces $\mathfrak{E}$, $\bar{\mathfrak{E}}$ with a
``biorthogonal'' basis $e_i\to\mathbf{e}_i=\mathbf{n}_i\otimes\mathbf{n}$,
$\bar{e}_j\to\bar{\mathbf{e}}_j=\bar{\mathbf{n}}_j\otimes\bar{\mathbf{n}}$
($\mathbf{n}$, $\bar{\mathbf{n}}$ are two-dimensional unit orthogonal
vectors, $\mathbf{n}_i$, $\bar{\mathbf{n}}_j$ are basis vectors of
spaces $\mathfrak{E}$, $\bar{\mathfrak{E}}$):
\begin{equation}\label{eq:2.3}
\bar{\mathbf{e}}_i\wedge\mathbf{e}_j
= (\bar{\mathbf{n}}_i,\mathbf{n}_j)\,\bar{\mathbf{n}}\wedge\mathbf{n}
= \delta_{ij}\,\bar{\mathbf{n}}\wedge\mathbf{n}
= \delta_{ij}\,\mathbf{b},
\end{equation}
where $\mathbf{b}$ is the unit bivector.  The vectors
\begin{equation*}
\mathbf{z}_j = \varphi_j\mathbf{e}_j + \bar{\psi}_j\bar{\mathbf{e}}_j,
\end{equation*}
where $\varphi_j$, $\bar{\psi}_j$ are complex numbers with
$\mathbf{n}_i^*=\bar{\mathbf{n}}_i$,
$\bar{\mathbf{n}}_i^*=\mathbf{n}_i$, belong to the space
$\mathfrak{E}\oplus\bar{\mathfrak{E}}$.

\noindent\textbf{Q6.}\ To the elementary event $\omega_j^i$ there is
assigned the measure $P$ of the bivector
\begin{equation}\label{eq:2.4}
\psi(\omega_j^i) = \tfrac{1}{2}\,\bar{\mathbf{z}}_j^*\wedge\mathbf{z}_j,
\end{equation}
i.e., $\psi(\omega_j^i)$ is $P(\psi(\omega_j^i))\ge 0$, where
$P(\mathbf{b})=1$.

According to Q3, if $A=\sum_j\omega_j^i$, then
$P(A)=\sum_j P(\psi(\omega_j^i))$.  The addition of axiom Q5 requires a
change in $\psi(\omega_j^i)$ in axiom Q2: now $\psi(\omega_j^i)$ is
determined by Q6.

It might seem that the introduction of the direct sum
$\mathfrak{E}\oplus\bar{\mathfrak{E}}$, as well as the product~\eqref{eq:2.4},
is excessive, and that it suffices to set $\bar{\psi}_j=-\bar{\varphi}_j$
in Q2.  Then one could get by with introducing a single axiom.

\noindent\textbf{QM5.}\ The sets of states $\{e_i\}$, $\{\bar{e}_j\}$
form complex vector spaces $\mathfrak{E}$, $\bar{\mathfrak{E}}$ with a
biorthogonal basis $e_i\to\mathbf{e}_i$, $\bar{e}_j\to\bar{\mathbf{e}}_j$,
$(\bar{\mathbf{e}}_i,\mathbf{e}_j)=\bar{\mathbf{e}}_i\cdot\mathbf{e}_j
=\delta_{ij}$, and if $\mathbf{z}=\sum_i\varphi_i\mathbf{e}_i$, then
$(\mathbf{z}_1,\mathbf{z}_2)=\sum_i\bar{\psi}_{1i}\psi_{2i}$.

The point is that non-relativistic QM is, though a very important, still
only a special case of quantum theory.  Relativistic QM and QFT remain
outside the picture: the condition $\psi(\omega_j^i)=|\psi_j|^2$ does not
hold either for the relativistic quantum theory of bosons, or for the
quantum theory of fermi-fields (see below).  The meaning of the
product~\eqref{eq:2.4} is explained in Section~4.

Let us give examples of vectors $\mathbf{z}_j$.

1.\ If $\bar{\psi}_i=-\bar{\varphi}_i$, then
$\psi(\omega_j^i)=|\psi_j|^2$ --- we obtain non-relativistic QM.

2.\ If $\bar{\psi}_i=\bar{\varphi}_i$, then
$\psi(\omega_j^i)=\tfrac{1}{2}(\psi_j\bar{\varphi}_j
-\bar{\varphi}_j\psi_j)$ --- we obtain the expression for the
``probability density'' in the quantum theory of relativistic bosons
(cf.\ question IV\,4).

In the case of fermions, axioms Q5, Q6 should be modified.

\noindent\textbf{Qf5.}\ In theories with anticommuting variables the sets
$\{e_i\}$, $\{\bar{e}_j\}$ form complex vector spaces $\mathfrak{E}_g$,
$\bar{\mathfrak{E}}_g$ with values in the Grassmann algebra and a
``biorthogonal'' basis $e_i^g=\mathbf{n}_i\theta\mathbf{n}$,
$\bar{e}_j^g=\bar{\mathbf{n}}_j\bar{\theta}\bar{\mathbf{n}}$
($\theta^2=\bar{\theta}^2=0$, $\theta\bar{\theta}=-\bar{\theta}\theta$,
$\bar{\theta}=\theta^*$; $\mathbf{n}$, $\bar{\mathbf{n}}$ are
two-dimensional complex orthonormalized vectors; $\mathbf{n}_i$,
$\bar{\mathbf{n}}_j$ are basis vectors of the spaces $\mathfrak{E}$,
$\bar{\mathfrak{E}}$):

The vector $\mathbf{z}_j$~\eqref{eq:2.3} is replaced by
\begin{equation}\label{eq:2.5}
\mathbf{z}_j^g = \psi_j\mathbf{e}_j^g
              + \bar{\psi}_j\bar{\mathbf{e}}_j^g,
\end{equation}
and the rule~\eqref{eq:2.4} is replaced by the following.

\noindent\textbf{Qf6.}\ To the elementary event $\omega_j^i$ there is
assigned the measure $P$ of the bivector
\begin{equation}\label{eq:2.6}
\psi_\theta(\omega_j^i) = \tfrac{1}{2}\,
\bar{\mathbf{z}}_j^{g*}\wedge\mathbf{z}_j^g,
\end{equation}
and $\psi_\theta(\omega_j^i)$ is
$P(\psi_\theta(\omega_j^i))\ge 0$, provided that
$P(i\bar{\theta}\theta\,\mathbf{b})=1$.

Setting $\bar{\psi}_j=\bar{\varphi}_j$, we have
$\psi_\theta(\omega_j^i)=P(i\bar{\theta}\theta\,\psi(\omega_j^i))
=|\psi_j|^2$.

Axioms Q5, Q6 complete the construction of the mathematical apparatus of
quantum theory.  In the case $\bar{\psi}_i=-\bar{\varphi}_i$ the scheme is
simplified, since $\psi(\omega_j^i)=|\psi_j|^2$ and a Hilbert space
$\mathcal{H}$ appears.  Instead of the vectors~\eqref{eq:2.3} one can
restrict oneself to vectors $\mathbf{z}=\sum_i\psi_i\mathbf{e}_i$ with
basis $\mathbf{e}_i\ne\bar{\mathbf{e}}_j$ and the multiplication rule:
\begin{equation*}
\mathbf{e}_i^*\cdot\mathbf{e}_j
= (\bar{\mathbf{n}}_i,\mathbf{n}_j)=\delta_{ij},\qquad
\mathbf{z}^*\cdot\mathbf{z}'
= (\bar{\boldsymbol{\psi}},\boldsymbol{\psi}')
= \sum_i\bar{\psi}_i\psi'_i,
\end{equation*}
i.e., in $\mathbf{e}_i$ one can suppress $\mathbf{n}$ (see Q5).  Let us
list the main features of the theory obtained in this way.

\medskip
\noindent I.\ \emph{Superposition principle for vectors:}
$\mathbf{z}=\sum_i\psi_i\mathbf{e}_i$ is a vector in $\mathcal{H}$, i.e.,
$\mathbf{z}=\psi'\mathbf{e}'$, $\mathbf{e}'\in\mathcal{H}$,
$\mathbf{e}'^*\cdot\mathbf{e}'=1$, where $\psi'$ is a complex number.  If
$\|\mathbf{z}\|=1$, then $|\psi'|=1$.

\noindent II.\ \emph{Theory of unitary transformations:}
$\sum_k U_{jk}^* U_{jk'}=\delta_{kk'}$; the vectors $\mathbf{e}_k'$ form
a new basis.

\noindent III.\ \emph{Operators:}
$\hat{A}=\sum_i A_i|i\rangle\langle i|$, where $|i\rangle=\mathbf{e}_i$,
$\mathbf{e}_i^*\cdot\mathbf{e}_j=\langle i|j\rangle$, $A_i$ are their
eigenvalues.

\noindent IV.\ \emph{A new class of elementary events:}
$\omega_j^i\to\langle i|j\rangle=U_{ij}$.

The last point means that quantum mechanics, while being included in
standard probability theory, is richer and, in a certain sense, more
substantive than the standard theory with non-negative numbers.

Noteworthy is the \emph{absence} of Planck's constant $h$ in the axioms.
This testifies to the fact that $h$ is not an indispensable attribute of
quantum theory.  The constant $h$ appears only in sufficiently complex
models with dimensional variables ($x$, $t$, etc.).  In Section~10.3 a model
of quantum theory with discrete space and time, not containing $h$, is
presented.

Axioms Q6, Qf5, Qf6 do not agree with the theorem of Section~2.1.  In
\eqref{eq:2.2} the discussion is about the square of the modulus of a
complex function $\psi(x)$, whereas in Q6 it is about the measure of a
bivector, and in Qf5, Qf6 about elements of the Grassmann algebra.  The
reason is that in axioms Q6, Qf5, Qf6 more complex structures are meant.
In the mentioned theorem time does not figure, to say nothing of Minkowski
space.  It is precisely the recourse to the bivector $\psi(\omega_j^i)$ in
Q6 that makes it possible to obtain the correct expression for
$\rho(x)=\frac{1}{2}(\psi^+\dot{\psi}^+-\dot{\psi}^+\psi^+)$ in
relativistic theory.  In the theorem there is no mention of Grassmann
variables, without which no even marginally substantive field theory can do.
Thus, the relative poverty of the content of the theorem~\cite{Lukacs} is
connected with the poverty of the classical probability theory taken as a
basis.  It does not consider the probabilistic schemes that arise in the
classical limit from modern QFT.

%======================================================================
\section{Quantum Mechanics and Quantum Field Theory}
\label{sec:QMQFT}
%======================================================================

The answers to the remaining questions of Group~II cannot be obtained within
the framework of QM\@.  For this it is necessary to pass to a more general
theory.  Such a theory is QFT\@.  In fact, it gives perfectly definite
answers to the remaining questions in the scope that was meant when they
were formulated in the period of the creation of QM.

\subsection{What is a particle?}
\label{sec:3.1}

Difficulties are caused already by the very concept of a particle in QM\@.
Essentially, one is dealing with the use of old ideas as applied to new,
hitherto unknown objects.  The peculiarity of the situation is that the new
objects also possess some properties of classical particles.  The problem is
resolved by QFT, according to which any particle is a quantum, a non-local
single-particle excitation of the corresponding field; this was specifically
pointed out, in particular, in the works~\cite{Blohintsev,Frenkel}.  Since
then, these works are usually not cited, since the assertion passed into the
category of the trivial and is disputed by no one.  Meanwhile, it is a key
assertion.  Surprisingly, this fact is ignored in all later discussions
devoted to the ``interpretation'' of QM\@.  As before, point particles and
non-relativistic theory are discussed.

And so, according to QFT, a particle is a non-local excitation of a field,
i.e., by no means a ``point''.  It is a structure formed by an ordered
collection of excited oscillators of the field.  Thereby all questions
connected with the idea of a particle as a point-like object are
automatically removed.  However, QFT, resolving one question, generates two
new ones.  If a particle is a non-local excitation, then:

1) How does one explain its ``point-likeness'', i.e., the possibility
of localization in a small region of space?

2) How does one explain its integrity, i.e., the fact that in interaction
processes a non-local excitation behaves as an indivisible object?

Finally, the question of the meaning of the wave function remains.  QFT
copes with these problems as well.

\subsection{What is the wave function?}
\label{sec:3.2}

The absence of an answer to the question ``What is the wave function?''
generates ``physical extremism''.  Thus, in~\cite{Fuchs} we read: ``\ldots
quantum theory does not describe physical reality.''  And further: ``\ldots
the wave function is not an objective entity''~\cite{Fuchs}.  Of course, in
such a case it cannot describe ``physical reality''~\cite{EPR}, which cannot
be subjective.  This underscores the importance of a clear understanding of
the essence of the wave function.

Quantum field theory gives a simple and clear answer to the question: what
physical reality corresponds to the wave function?  If a particle is a
single-particle excitation of a field, then its wave function is a function
describing this excitation.  Like any field configuration, a single-particle
excitation is described by a function of coordinates.  The relationship
between the wave function and space thereby ceases to be something
mysterious: a single-particle excitation of a field has a perfectly clear
spatial structure, since the field itself lives in space.

Thus, we have:
\begin{equation*}
\text{particle} = \text{field excitation},\qquad
\text{wave function} = \text{function of excitation of a field.}
\end{equation*}

Let us also indicate the equation that the single-particle wave function
satisfies.  Let $\Phi(x_1,\ldots,x_N)$ be the wave function of the field
$\varphi(x)$, where $x_i$ and $N$ denote the occupation numbers.  For the
free field $\Phi$ satisfies equations that are derived from the field
Lagrangian.  The single-particle wave function
$\Phi_1(x)=\langle 0|\varphi(x)|\mathbf{k}\rangle$ satisfies the free
field equation, e.g., the Klein--Fock--Gordon equation for a scalar field,
the Dirac equation for a spinor field, and so on.  The non-relativistic
limit gives the Schr\"odinger equation.  It is well known that
$\Phi_1(x)\ne\delta^{(3)}(\mathbf{x}-\mathbf{x}_0)$, i.e., the excitation
is always non-local, though possibly small.\footnote{It was specifically
pointed out, in particular, in papers~\cite{Blohintsev,Frenkel}.}

\subsection{Point-likeness and integrity of particles}
\label{sec:3.3}

Once a particle is identified with a field excitation, the questions of
``point-likeness'' and integrity are resolved by the mathematical properties
of field configurations.

Since in the Hamiltonian of any bosonic field $\varphi$ there is a term
proportional to $(\nabla\varphi)^2$, a discontinuity of $\varphi$ means
the appearance of a non-integrable singularity $(\nabla\varphi)^2$ in the
Hamiltonian, i.e., a field configuration of infinite energy.  It follows
that upon interaction with another field, an excitation of a given field is
either absorbed entirely, or entirely (and as a whole) scattered.

The above considerations generalize in an obvious way to the process of
scattering of particles as well.  This explains the \emph{integrity} of
particles, i.e., the fact that non-local field excitations behave as
indivisible objects.  This also explains their ``point-likeness'' ---
non-local excitations interact as point-like objects.

What does the possibility of localizing a particle mean from the standpoint
of QFT\@?  Only that, in some region of space where the field is excited, it
can interact with another particle, i.e., with another (or the same) field.
The act of interaction always takes place at a point.  This is precisely the
act of registration of a particle --- at any rate, its most essential
element.  Since particles act as integral non-local objects, when indicating
their position one can speak only about the region of space in which the
field is excited.  In this way the very concept of the coordinate of a
micro-object is clarified.

The region of single-particle excitation of a field cannot be much smaller
than the Compton wavelength $\lambda_c=\hbar/mc$ --- otherwise new
particles would be born and the state would cease to be single-particle.
Thus, a particle can be \emph{registered} in a practically arbitrarily
small region (the minimal size is determined by the Planck length $l_P$),
but \emph{localized} in a region of minimal size $\sim\lambda_c$.

The wave function in QM depends on the coordinate, i.e., on a parameter
that in no way characterizes the particle; it characterizes the particle's
position --- with its help the region of space
$\Omega=\{x\in\Omega\mid\psi(x)\ne 0\}$ in which the particle can be
detected is fixed.  Such a mode of description is acceptable in
non-relativistic physics, at low energies, when the de~Broglie wavelength is
much greater than the Compton wavelength $\lambda_c$.  Then, approximately,
$|\psi(x)|^2$ is the probability density of finding the particle ``at the
point~$x$''.  In relativistic physics such an interpretation loses its
meaning --- both for a formal reason (the density
$i\varphi^+\!\overset{\leftrightarrow}{\partial_0}\varphi^+$ is not
positive definite) and in substance (any quantum of a field is an
intrinsically non-local object).  The second circumstance justifies the
first.

The practice of applying QFT to experiment shows that the function
$\Phi_1(x)=\langle 0|\varphi(x)|\mathbf{k}\rangle$ is indeed the
coordinate representation of the state vector of the photon.  But it does
not allow one to obtain a local probability density for finding a
relativistic particle at a point --- such information is devoid of meaning.
Consequently, attempts at ``interpretation'' of QM that rest on the
assumption of strict point-likeness of particles are also devoid of meaning.
When elucidating the nature of the quantum description of the microworld,
one cannot ignore such a fundamental fact as the spatial extent of field
quanta.

Let us dwell on one more important question.  The equation of motion of a
free field is linear, and therefore the identification of the function
describing a field excitation with the wave function raises no objections.
The question arises when interaction is switched on.  Now the evolution of
the field is described by a nonlinear equation, and the identification is
impossible.  But the nonlinearity of QFT means the possibility of changing
the number of particles, which presupposes the transition to relativistic
energies.  This requires a generalization of the concept of the wave
function.  In the linear case its role was played by a dynamical variable
--- the field $\varphi(x)$ itself (in~\cite{Prokhorov1999,Prokhorov2000}
the possibility of the existence of such probabilistic schemes was shown).
In QFT the role of the wave function is played by a functional of the
dynamical variable --- the Fock functional; for it, both the classical and
quantum equations of motion are linear.

Finally, there is yet another circumstance requiring clarification.
Classical fermionic fields are described by functions with values in the
Grassmann algebra.  Can one, in this case, identify the field with the wave
function?  Axioms~Qf5, Qf6 of Section~2 give an affirmative answer to
this question.

\subsection{Wave--particle duality and the uncertainty relations}
\label{sec:3.4}

The question of how the ``wave'' (wave function) and the ``particle''
(material point) are related to each other has stood from the very first
days of the appearance of quantum mechanics.  Schr\"odinger initially
supposed that a particle is a wave packet.  But a wave packet disperses
with time, while a particle does not.  De~Broglie proposed that linear
quantum mechanics is the limit of some nonlinear theory admitting solutions
with singularities, which were then identified with particles (the theory
of the double solution).  But linear quantum mechanics describes
experiment with enormous precision.  The weak point of these and subsequent
attempts at ``interpretation'' of quantum mechanics was that the problem was
considered in its original formulation, in the form in which it stood in
the 1920s, i.e., before the creation of modern quantum field theory.

In fact, the dilemma was resolved already in 1927, when P.\,A.\,M.~Dirac
\cite{Dirac1927}, P.~Jordan, O.~Klein~\cite{JordanKlein} (see
also~\cite{JordanWigner}) introduced the procedure of ``second
quantization'' (they ``quantized'' the wave function).  ``The method of
second quantization \ldots was regarded by all as a mathematical expression
of wave--particle duality, i.e., an expression of the fact that `the
corpuscular picture and the wave picture are different aspects of one and
the same physical reality'.  A better agreement between formalism and
interpretation would have been hard to
desire''~\cite{Jammer}.

Indeed, if a particle is an excitation of a field, then it can manifest
itself in different ways depending on the circumstances.  On the one hand,
by virtue of the continuity of the field, it behaves as a corpuscle
(``integrity'' of particles); on the other hand, one can observe an
interference pattern for a field excitation with a sufficiently narrow
energy (frequency) distribution.  The wave \emph{is} the ``particle'', and
the particle \emph{is} the wave.

Equally simply is the essence of the uncertainty relations elucidated.  The
act of registration of a particle (excitation) begins with an act of
interaction of one of the excited oscillators of the given field with some
other field (for example, the electromagnetic field with the
electron--positron field).  It is clear that any one of the activated
oscillators can begin to interact; therefore one cannot speak of the
``point'' at which this single-particle excitation is located.  One can
speak only of the size of the region in which the field is excited, i.e.,
of $\Delta x$.  The distribution over momentum, inherent to a given
excitation, is obtained with the help of the Fourier transform, which is
also well known from classical physics.  The rest is a matter of
mathematics: from Fourier analysis it is known that
$\Delta x\,\Delta k\ge 1$, where $k$ is the wave number ($2\pi/\lambda$).
In QM $p=\hbar k$, whence the inequality $\Delta x\,\Delta p\ge\hbar$
follows (if $\Delta x$ and $\Delta p$ are root-mean-square deviations from
the mean values, then in the inequality one should replace
$\hbar\to\hbar/2$).  Thus, the uncertainty relations characterize the
properties of a sufficiently complex non-local object --- a single-particle
excitation of a field.  The question is exhausted.

\subsection{On the completeness of the quantum description}
\label{sec:3.5}

\emph{Definition}: ``\ldots a theory is incomplete if within the framework
of the theory one can devise [and carry out] an experiment whose
performance in itself does not contradict the given theory, but whose
result the theory is unable to predict''~\cite{Markov}.

Obviously, QM does not satisfy the given criterion of completeness.  One
can devise and carry out an experiment to determine the position of a
particle after an act of scattering.  But QM cannot reliably predict its
result.  The question is: is it impossible to have more complete
information about the state of the particle than the uncertainty relations
allow?  Perhaps there exist additional characteristics of the particle, as
yet unknown to us (``hidden parameters''), which would allow one, for
example, to indicate precisely its coordinate and momentum?  The problem of
the existence of such parameters in QM was first investigated by J.~von
Neumann~\cite{vonNeumann}.  Subsequently this direction received
considerable development, mainly after the
works~\cite{EPR,Bell1,Bell2}.

A different definition of completeness seems more expedient.  A theory is
incomplete if there exists additional, unaccounted-for information about
the objects and processes that it describes.  Strictly speaking, no closed
theory (i.e., a theory specified by a finite number of axioms) can be
complete.  For example, classical mechanics of material points is
incomplete, because any body possessing mass has a finite size and
structure; classical theory is incomplete, because any body and field obey
the laws of QM; and so on.  The question of the completeness of QM arose,
firstly, in connection with the uncertainty relations (can they be
violated?), and secondly, in connection with the probabilistic nature of QM
(can one return to determinism?).  QFT answers both questions in the
negative, since a ``particle'' is a single-particle excitation of a field
(Section~3.1), and quantum theory by its nature is stochastic
(Sections~2,~4,~5).  Incidentally, in the case of a particle on a circle,
the uncertainty relations are trivialized:
$\Delta p_\varphi\,\Delta\varphi\ge 0$, where $\Delta p_\varphi$ and
$\Delta\varphi$ are root-mean-square deviations from the mean values of
momentum and angle; both of them, for example, in a state with definite
momentum, satisfy $\Delta p_\varphi=0$, $\Delta\varphi=2\pi$.  The
uncertainty relations formally do not differ from those of classical
theory, although the mechanics remains quantum.  However, there are also
more subtle questions.

In~\cite{EPR} the discussion concerned the quantum description of two
particles, namely, the influence of experiments on one particle on the
state of another, distant particle (see Section~10.2).  It was shown that
according to the rules of QM (i.e., of the non-relativistic theory of
systems with a finite and conserved number of particles), the experimenter,
by ascertaining the state of the first particle, instantaneously
predetermines the state of the second (even if the operators representing
the observables do not commute).  In essence, the discussion here is about
the problem of combining the principle of locality (or causality in the
sense of the special theory of relativity) with non-relativistic QM, i.e.,
about the physical meaning of the wave function.

J.\,S.~Bell posed the question~\cite{Bell1,Bell2}: what would happen if
the state of a particle, within the framework of the
approach~\cite{EPR}, could be characterized in more detail?  It was
assumed that the theory would remain probabilistic.  It turns out that in
such a case certain inequalities must hold.  They contradict
experiment~\cite{Tittel,Weihs}.

Let us dwell on the main features of the cited works.  First of all,
concerning the character of the ``hidden parameters''.  Suppose the nuclei
of oxygen ${}^{16}$O and ${}^{18}$O are being studied.  At low energies and
with not too great precision of the experiments, they are
indistinguishable and are described identically.  But they have a hidden
parameter --- the baryon number (16 and 18, respectively).  It is clear
that the existence of such hidden parameters is possible, but their
discovery does not solve the problems of the quantum description.  If there
existed parameters allowing one to refine the position of a particle with a
definite momentum, then this, of course, would change the description of
micro-objects and would require a refinement of QM\@.  The
works~\cite{Bell1,Bell2} give an answer to the question of how this fact
would manifest itself in experiment.  Let us stress that in
fact~\cite{Bell1,Bell2} compare the predictions of quantum theory with
those of a certain modified theory, which is not explicitly formulated and
whose self-consistency is not proven.  Consequently, Bell was considering
the consequences of some compromise theory, intermediate between the
classical and the quantum.

Meanwhile, the answer to the question of the completeness of QM is obvious:
quantum mechanics is incomplete, but not in the sense that was meant in the
period of its formation.  It is called upon to describe individual
particles, pairs of interacting particles, and so on.  It can describe a
photon; it can describe an electron; but it cannot describe their
interaction.  The essence of the interaction of a photon with an electron
reduces to its absorption or emission by the latter.  But for the
description of such processes a theory of dynamical systems with a variable
number of particles is required, i.e., quantum electrodynamics.  What has
been said applies to any particles: QM is merely a sector of QFT, which
only approximately can be considered independent.  Usually one speaks of QM
as the non-relativistic limit of QFT --- however, non-relativistic photons
do not exist.  Thus, QM is incomplete, but not because hidden parameters
exist, but because, being a part of a more general theory, it is not in a
position to describe even the simplest systems of two particles --- an
electron and a photon.  At first glance this may seem like a special case,
not essential for solving the problem.  In reality, it is one of the key
points, important for understanding the essence of the quantum description.

Another reason for the incompleteness of QM is relativism.  Taking it into
account makes unfit the interpretation of the wave function in the
coordinate representation, accepted in non-relativistic QM\@.  The
probability density in $x$-space, allowed by the theory of relativity, is
not positive definite; but a particle (a quantum of a field) is not a
point, either --- i.e., in this respect, too, a transition to a more
general theory (QFT) is necessary.

Thereby the problem is transferred to a higher level and reduces to the
question: ``Is QFT complete?'' --- a question which, apparently, has not
been posed.  However, the transition to QFT is extremely important in
another respect as well --- it generates new questions (for example, about
the possibility of splitting quanta), answers to which may help to sort out
certain paradoxes of QM.

\subsection{Paradoxical states and superselection rules.  Identity of
particles}
\label{sec:3.6}

\medskip
\noindent\textbf{Paradoxical states.}

If a particle is a field excitation, then what can one say about a state in
which the field is excited in two non-overlapping regions of space?  Is
this a single-particle or a two-particle state?

Let us turn to the single-particle wave function.  Suppose that
\begin{equation*}
f_n(x)=f_1(x)+f_2(x),\tag{3.5}
\end{equation*}
where the functions $f_1$ and $f_2$ are non-zero in regions $\Omega_1$
and $\Omega_2$, $\Omega_1\cap\Omega_2=\varnothing$, i.e.,
$f_1(x)\,f_2(x)=0$.  From the standpoint of QFT, the vector
\begin{equation*}
A^{(+)}(f_n)|0\rangle\tag{3.6}
\end{equation*}
is a single-particle state, since it is an eigenvector of the
particle-number operator $\hat{N}$ with eigenvalue~1.  We shall call such
states \emph{paradoxical} or \emph{exotic}.  Their exotic character lies in
the fact that they consist of two physically disconnected parts.  Such wave
functions are allowed by the theory of Hilbert spaces, but have apparently
not been discussed until now, in spite of their importance for understanding
the essence of the Einstein--Podolsky--Rosen (EPR) paradox~\cite{EPR}.
Indeed, $|f_n|^2=|f_1|^2+|f_2|^2$, so the particle can be detected
either in the region $\Omega_1$ or in the region $\Omega_2$.  According to
the postulates of QM, detection of the particle in the region $\Omega_1$
excludes its presence in $\Omega_2$ --- a ``reduction'' of the wave
function takes place.  But the regions $\Omega_1$, $\Omega_2$ are separated
by a space-like interval; therefore, according to QFT, an excitation in
$\Omega_1$ cannot influence an excitation in $\Omega_2$, i.e., absorption
of the excitation in $\Omega_1$ by another field does not affect the
excitation in $\Omega_2$ (the regions $\Omega_1$, $\Omega_2$ may be
located in different galaxies).  In essence, this is the key to the EPR
paradox.  QM prescribes instantaneous reduction of the wave function, while
QFT forbids it: spatially separated regions are not causally connected, and
therefore instantaneous transfer of information from one region to another
is impossible.

Thus, QFT introduces very serious corrections into the orthodox treatment
of QM\@.  At the same time, new possibilities are discovered.  For example,
suppose there is an ordinary single-particle excitation characterized by a
function $f$ with a connected compact support.  What will happen if it is
converted into an exotic one, i.e., if one splits the field excitation in a
connected region into excitations in two disconnected regions?  At first
sight this is impossible, by virtue of the requirement of continuity of
the field and the integrity property of particles (see Section~3.3).
However, for example, a photon has a finite size --- the coherence length,
which can be large even by macroscopic standards --- and discontinuous
fields are forbidden only in the continuous limit.  For discrete structures
(such as a chain of oscillators), the energy of a discontinuous field
configuration $\varphi$ is estimated as
$\sim\Delta x\,(\Delta\varphi/\Delta x)^2=(\Delta\varphi)^2/\Delta x$,
and if $\Delta x\sim l_P$, then discontinuous configurations of the field
are admissible but possess very large energy (of order
$(\Delta\varphi)^2 M_P$, $M_P\approx 2\times 10^{-5}$~g).  This, on the
one hand, explains the integrity of quanta in the microworld at energies
$E<10^{19}$~GeV, and on the other hand opens the possibility of splitting
quanta in processes involving macroscopic bodies, which in the scales of
the microworld possess inexhaustible reserves of energy.

What properties would the fragments of split quanta possess?  For
concreteness, let us take an electron (a quantum) split into two halves.
Each half is also an excitation of the electron--positron field, and
therefore all the attributes of the field inherent to it (mass, spin)
remain unchanged, but their wave functions will be normalized to $1/2$.
The electric charge of the electron is the amplitude of the probability of
emitting a photon; therefore the electric charge of each half will equal
$e/2$.  It is clear that if quanta can be split, then they can also be
merged, i.e., one can expect the existence of quanta of the
electron--positron field with charges $2e$, $3e$, etc.  Experimental
consequences of this possibility are discussed in Section~7.

The presence of field states~(3.6) with functions~(3.5) raises the question
of their essence.  Consider the state
\begin{equation*}
A^{(+)}(f_1)\,A^{(+)}(f_2)|0\rangle,\tag{3.7}
\end{equation*}
where $A^{(+)}$ contains only creation operators.  Formally, this is a
state of two non-interacting particles with wave functions $f_1$, $f_2$.
And it is such in substance as well.  What, then, is the difference
between states~(3.6) and~(3.7)?  After all, in both cases the field is
excited in the same regions.  The difference is that (3.6) is a
\emph{superposition} of two excitations of the field in disconnected regions
of space.  It is precisely such states that generate paradoxical
situations.  Field excitations are real, objectively existing formations
possessing all the attributes of material particles (energy, momentum,
etc.).  The inclusion in the theory of state vectors~(3.6), allowed by QM,
leads to physically unacceptable consequences: registration of the
excitation in the region $\Omega_1$ annihilates the excitation in the
causally disconnected region $\Omega_2$.  Of course, such a situation is
inadmissible under the condition of the reality of the field as a physical
object.

Where is the way out?  The simplest --- forbid the superposition of the
states $A^{(+)}(f_1)|0\rangle$ and $A^{(+)}(f_2)|0\rangle$ with
functions $f_1$, $f_2$~(3.5).  Are there, in quantum theories, prohibitions
on the superposition of certain state vectors?  Yes!  These are
superselection rules~\cite{WWW} (for more detail, see Section~10.1).
If, for example, $\psi_b$ and $\psi_f$ are, respectively, the wave
functions of a boson and a fermion, then the state
$\psi=c_1\psi_b+c_2\psi_f$ is forbidden by a superselection rule.  The
Hilbert-space vector (ray) $\psi$ does not ``go over into itself'' under a
rotation of the coordinate system through the angle $2\pi$ (at this
$\psi_b\to\psi_b$, $\psi_f\to-\psi_f$), which is unacceptable, since
rotation by $2\pi$ is the identity transformation, and all vectors should
go over into themselves.  Here we have an example of a restriction imposed
by the properties of space on the rules of QM\@.  Another example --- the
prohibition on superposition of states with different electric charges
$e_1$, $e_2$: $\psi=c_1\psi_{e_1}+c_2\psi_{e_2}$.  This prohibition was
originally regarded as an empirical rule.  Later it was
clarified~\cite{Prokhorov1990} that it follows from the rules of
quantization of the electromagnetic field as a system with first-class
constraints.  In this case, the peculiarities of the dynamics of the field
already forbid certain states that were initially considered admissible.
A characteristic sign of the appearance of such states is the existence of
operators commuting with all physical operators of the theory (example: the
operator of electric charge~$Q$); this is equivalent to the absence of
operators commuting with all physical operators and transforming the
vectors $\psi_{e_1}$, $\psi_{e_2}$ into each other.  In relation to the
state~(3.5), it is clear that within QM and QFT there do not exist such
physical operators (the operation consists of the instantaneous transfer of
an excitation from the region $\Omega_1$ to $\Omega_2$).  An operator
transforming the normal state into an exotic one would also be acceptable.
It could be associated with the action of a macroscopic body.  But the
introduction of macroscopic bodies into the theory means a transition to
another, unitarily inequivalent Hilbert space, i.e., a transition to a
different theory, since all vectors of the original space are orthogonal
to the vectors of the final one describing the macroscopic body as well.
Transitions between states belonging to unitarily inequivalent spaces are
admissible in the physics of phase transitions, due to the
non-separability of the Hilbert space of QFT, but there they are not
associated with the artificial introduction of extraneous bodies into the
theory.

\medskip
\noindent\textbf{Identity of particles.}

The identification of particles with field excitations sheds light on the
nature of the \emph{identity of particles}.  The term itself requires
clarification.  In the commonly accepted sense (and in classical physics)
it means ``the same'', ``alike'', ``completely coincident'', ``identical'',
and does not have a precise meaning.  For example, two pairs of
factory-made shoes of the same size are alike and quite ``coincident''.
But, obviously, they differ in microscopic details --- if only because of
inaccuracies in cutting.  Physics gives a more precise criterion of
\emph{statistical indistinguishability}, which leads to a tangible
difference when computing the probabilities of events (classical and
quantum statistics).  On the other hand, the contents of some document and
its photocopy are identical, and in a hypothetical theory operating with
the statistics of the contents of such documents, they would be identical,
i.e., statistically indistinguishable.

In quantum physics the idea of identity of particles manifests itself,
firstly, in the property of symmetry or antisymmetry of wave functions of
two or more ``identical'' particles (electrons, protons, etc.), and
secondly, in the appearance of quantum statistics (the statistical weight
of $N$ particles decreases by a factor of $N!$ --- a consequence of the
fact that a configuration with permuted particles is not counted as a new
possibility when computing the probabilities of events).

A literal understanding of the symmetry properties of the wave function
leads to paradoxical situations.  Consider the state of two distant
electrons (let one be on Earth and the other in a neighboring galaxy) with
normalized wave functions $f_1(x_1)$, $f_2(x_2)$ as in~(3.5), i.e.,
$f_1(x)\,f_2(x)=0$, $\|f_1\|=\|f_2\|=1$.  In the singlet state, their
normalized wave function is
\begin{equation*}
\Psi(x_1,x_2)=\tfrac{1}{\sqrt{2}}\bigl[f_1(x_1)f_2(x_2)-f_2(x_1)f_1(x_2)
\bigr].
\end{equation*}
In experiments with the electron on Earth, one should average over all
states of the extra-terrestrial electron.  Then the probability density for
the first electron is given by the integral
\begin{equation*}
\int dx_2\,|\Psi(x_1,x_2)|^2 = \tfrac{1}{2}\bigl[|f_1(x_1)|^2+|f_2(x_1)|^2
\bigr].\tag{3.9}
\end{equation*}
But on Earth $f_2(x_1)=0$, so that a blind adherence by the experimenter to
the rule of symmetrization of the wave function of all electrons leads to
the absurd: the probability distribution turns out to be incorrectly
normalized.  Meanwhile, in statistical computations one must integrate over
$x_1$, $x_2$ in the \emph{entire space}, and then the paradox does not
arise.  Thus, one should speak of the \emph{statistical
indistinguishability} of particles.

Quantum field theory clarifies the question.  If particles are
single-particle excitations of fields, then, of course, one cannot say that
it is impossible to distinguish one of the spatially separated excitations
of a field from another.  For example, let us take two photons.  Suppose
the coherence length of one of them is 1~cm, and that of the other is
1~km.  The assertion of their indistinguishability is devoid of meaning ---
distinguishing them presents no difficulty; confusing them is impossible.
In a statistical treatment, however, one should take into account all
possibilities (integrate over the entire phase space), and then the
statistical indistinguishability of field excitations does not look so
surprising, since one is dealing with excitations of one and the same field
oscillators.

Let us pass to the questions of Group~I. The problem of the nature of the
wave function is resolved within the framework of QFT.  But this does not
resolve the main problem of quantum theories --- the appearance of
probability amplitudes in mechanics, since QFT itself is based precisely on
them.  The question of the nature of Planck's constant $h$ also remains
without an answer.  It turns out that it is precisely the constant $h$ that
serves as the key to unravelling the nature of the quantum description.

%======================================================================
\section{Planck's Constant $h$}
\label{sec:Planck}
%======================================================================

\subsection{Theories with a universal constant of the dimension of action}
\label{sec:4.1}

We shall call a dimensional constant \emph{universal} if it sets the scale
in all problems of the theory, regardless of the nature and number of
degrees of freedom of a particular system.  In classical non-relativistic
mechanics there are no universal constants, neither dimensional nor
dimensionless.  Let us ask the question: does there exist a classical
mechanics of a particle in which a fundamental constant of the dimension of
action is present?  The dimension of action is shared by: the action itself,
angular momentum, and the volume (area) of PS\@.  Action is not conserved in
the course of motion; angular momentum has meaning only in spaces of
dimension $n\ge 2$; therefore the third possibility remains.  Let us dwell
on it in more detail.

\medskip
\noindent\textbf{Hamiltonian mechanics.  Compact phase space.}

To specify the Hamiltonian mechanics of a system, one must specify the PS,
a symplectic structure on it (or Poisson brackets), and the Hamiltonian.
The volume of PS is conserved during motion (Liouville's theorem), and
therefore a theory with a compact phase space is admissible.  This is
precisely the classical Hamiltonian mechanics with a fundamental constant
having the dimension of action.  Let us consider the simplest example of
such a mechanics.

Let the phase space of a particle be the sphere $S^2$ of radius $R$.  Then
the symplectic 2-form of area is
$\omega_2=R^2\sin\theta\,d\varphi\wedge d\theta
=\omega^{-1}(\varphi,\theta)\,d\varphi\wedge d\theta$,
where $\varphi$, $\theta$ are spherical angles, and the Poisson bracket is
\begin{equation*}
\{f(\varphi,\theta),g(\varphi,\theta)\}
= \omega(\varphi,\theta)\left(
  \frac{\partial f}{\partial\varphi}\frac{\partial g}{\partial\theta}
- \frac{\partial f}{\partial\theta}\frac{\partial g}{\partial\varphi}
  \right).
\end{equation*}
The area of the surface of the sphere (the ``volume'' of PS) is
$\int\omega_2=4\pi R^2$.  Having specified some Hamiltonian
$H(\varphi,\theta)$, we obtain an example of classical Hamiltonian mechanics
with a fundamental constant having the dimension of action ($4\pi R^2$).

By itself this example may be curious, but says little.  Its essential
feature is the possibility of introducing in a natural way a probabilistic
distribution on PS.  As is known, the unique invariant distribution on the
sphere is given by the density $\rho(\varphi,\theta)=1/4\pi R^2$.  This
fact, though noteworthy, also gives little by itself.  In order to reveal
the physical content of the theory, one must represent it in some other
form.  For this we use a suitable non-canonical transformation.

\medskip
\noindent\textbf{Non-canonical transformations.  Projection of the sphere
onto the complex plane.}

Non-canonical transformations in Hamiltonian mechanics, unlike canonical
ones, are not so well known, but are equally important.  Since they reduce
to a certain change of canonical variables, the volume of PS does not change
under them, and their use is admissible.  One such transformation is well
known and widely used --- the transition from standard canonical variables
$q$, $p$ to complex variables $z$, $\bar{z}$: $z=(q+ip)/\sqrt{2}$,
$\bar{z}=(q-ip)/\sqrt{2}$.  They have different Poisson brackets:
$\{q,p\}=1$, $\{z,\bar{z}\}=-i$.  The transition $q,p\to z,\bar{z}$ is
used in quantum theory (oscillator in the Dirac representation): operators
$z$, $z^+$ are identical to the annihilation and creation operators $a$,
$a^+$.

If a Hamiltonian mechanics is formulated on the sphere, then another example
of a non-canonical transformation is the ordinary stereographic projection
of the sphere onto the complex plane $Z$ (Riemann sphere).  It is defined
by the transformation $\varphi,\theta\to Z$, $|Z|=2R\cot(\theta/2)$,
$\arg Z=\varphi$.  The symplectic 2-form is given by
\begin{equation*}
\omega_2 = \frac{i}{2}\left(1+\frac{Z\bar{Z}}{4R^2}\right)^{-2}
dZ\wedge d\bar{Z}
= \omega^{-1}(Z,\bar{Z})\,dZ\wedge d\bar{Z},
\end{equation*}
whence follows the expression for the Poisson bracket
\begin{equation*}
\{f(Z,\bar{Z}),g(Z,\bar{Z})\}
= \omega(Z,\bar{Z})\left(
  \frac{\partial f}{\partial Z}\frac{\partial g}{\partial\bar{Z}}
- \frac{\partial f}{\partial\bar{Z}}\frac{\partial g}{\partial Z}
  \right).
\end{equation*}
The transition to the Riemann sphere also does little to clarify the
peculiarities of a theory with a compact PS.

The following mapping of the sphere onto the complex plane admits a
visualizable physical interpretation:
\begin{equation}\label{eq:4.1}
|z|^2 = \frac{\ln 2}{\beta}\cdot 3R^2\sin^2(\theta/2),\qquad
\arg z = \varphi,
\end{equation}
\begin{equation}\label{eq:4.2}
\omega_2 = \frac{i}{2}\,dz\wedge d\bar{z}\;e^{-\beta z\bar{z}}.
\end{equation}

Setting in~\eqref{eq:4.2} $z=(q+ip)/\sqrt{2}$, $1/\beta=kT$, we obtain
the Gibbs distribution for the oscillator:
\begin{equation}\label{eq:4.3}
\omega_2 = e^{-H/kT}\,dq\wedge dp,
\end{equation}
where $H=(p^2+q^2)/2$.  Thus, a dynamical system with one degree of
freedom and a spherical phase space is equivalent to an oscillator in a
thermostat.  Passing to a new Hamiltonian
$H\to H=p^2/2m+m\omega^2 q^2/2$ and equating the area to $h$, we have
\begin{equation}\label{eq:4.4}
\beta\omega = \frac{1}{\hbar}\qquad (\hbar = h/2\pi).
\end{equation}

The mean energy of the oscillator equals
\begin{equation}\label{eq:4.5}
\bar{E} = h^{-1}\int dq\,dp\;H\,e^{-\beta H} = \frac{1}{\beta},
\end{equation}
i.e., taking into account~\eqref{eq:4.4},
\begin{equation}\label{eq:4.6}
\bar{E} = \hbar\omega.
\end{equation}

The constant $h$ is naturally identified with Planck's constant.  Taking
into account, further, that $e^{-\beta H}\,dq\,dp/h$, on the one hand,
specifies the probability distribution on PS, and on the other hand is
proportional to the area element of PS, we have for the probability
corresponding to the event $A$:
\begin{equation}\label{eq:4.7}
P(A) = \int_A e^{-\beta H}\,\frac{dq\,dp}{h} = \frac{S(A)}{h},
\end{equation}
where $S(A)$ is the area of the PS region $A$.

\subsection{Random variables and Fock space}
\label{sec:4.2}

As was noted, a system with a compact phase space $\Gamma$ (here $\Gamma$ is
a sphere) opens the way to a probabilistic description.  On the sphere there
is a natural measure (area) and a probability space
$(\Omega,\mathcal{A},P)$~\cite{Kolmogorov,Loeve} appears naturally:
$\Omega$ (the set of elementary events) is the set of points of the sphere,
$\mathcal{A}$ is the Borel set (algebra) of all subsets of $\Omega$, $P$ is
the non-negative function (probability distribution) defined on
$\mathcal{A}$; we define it by the condition that the probability density
$\rho$ on the sphere is constant:
$\rho(\varphi,\theta)=1/4\pi R^2$.

Even more important is the fact established above: the sphere admits a
mapping onto the (complex) plane such that the system ``particle on a
sphere'' can be interpreted as the system ``harmonic oscillator in a
thermostat''.  This circumstance introduces fundamentally new elements into
the theory.  An oscillator whose state is characterized by a Gibbs
distribution with some temperature is an object described within the
framework of probability theory.  But in this case a temperature appears,
and the possibility of dissipative processes, where the temperature is
expressed through $h$ and is a world constant.  Even more remarkable is the
fact that the complex random variables form a Fock space.

Let us discuss the matter in more detail.  In a one-dimensional space
$\mathbb{R}_1$ one can propose three different constructions: classical
Hamiltonian mechanics, statistical mechanics with a flat PS, and quantum
mechanics.  One need only specify the Hamiltonian (and the temperature in
the second case; the Poisson bracket is taken to be standard, since PS is a
plane).

In the case of a \emph{spherical phase space}, the Poisson bracket is
determined by the measure of area of the sphere, i.e., by the non-trivial
symplectic form.  In this case it turns out that: 1) the problem of a
``free'' particle is equivalent to the problem of a particle with PS
$\mathbb{R}^2$ with a non-trivial symplectic form; 2) upon transition to
the statistical description we have the system ``harmonic oscillator in a
thermostat''; 3) the orthogonal random variables of the system form a basis
in Fock space.  The first two assertions have been demonstrated above.  Let
us prove the third.

Denote
\begin{equation}\label{eq:4.8}
d\mu(z,\bar{z}) = \frac{i}{2\pi}\,dz\wedge d\bar{z}\;e^{-z\bar{z}}
= \frac{1}{2\pi}\,dq\wedge dp\;e^{-(q^2+p^2)/2}
\qquad (z=\tfrac{q+ip}{\sqrt{2}});
\end{equation}
the measure~\eqref{eq:4.8} is normalized to unity.  Entire complex
functions $f(z)$ of growth order $\rho\le 2$ form the set of complex random
variables~\cite{Loeve}, and
$\langle|f(z)|^2\rangle=\int d\mu\,|f(z)|^2$.  They form a Hilbert
space with the scalar product~\cite{Loeve}:
\begin{equation}\label{eq:4.9}
(f,g) = \int d\mu(z,\bar{z})\;f(z)\,g^*(\bar{z})
\qquad (g^*(\bar{z})=\overline{g(z)}).
\end{equation}
The Hilbert space with scalar product~\eqref{eq:4.9} is Fock
space~\cite{Harte} (a ``scalar field in space-time $(0+1)$'' --- from the
field only one oscillator remains).  Indeed, as is easily verified, the
functions $Z_n(z)=z^n/\sqrt{n!}$ form an orthonormal basis in this space
(see Section~10.4):
\begin{equation*}
(Z_n,Z_m) = \int d\mu\; Z_n(z)\,Z_m^*(\bar{z}) = \delta_{nm}.
\end{equation*}

Further we have:
\begin{equation}\label{eq:4.10}
z\,Z_n(z) = \sqrt{n+1}\;Z_{n+1}(z),\qquad
\frac{d}{dz}Z_n(z) = \sqrt{n}\;Z_{n-1}(z),
\end{equation}
i.e., multiplication by $z$ and differentiation with respect to $z$ of the
function $Z_n(z)$ are identical to the action of the operators $a^+$ and $a$
on the state $|n\rangle$ of the harmonic oscillator.  The fact that the
operator $a=d/dz$ is conjugate (by the scalar product formula) to the
multiplication operator by $z$ (i.e., $a^+$) follows from the definition of
the scalar product~\eqref{eq:4.9} (see Section~10.4), and therefore the
notation $Z_n(z)=(z|n)$ is appropriate.  The complex random variables
$Z_n(z)$ are eigenfunctions of the energy operator of the harmonic
oscillator $\hat{H}=a^+a+\tfrac{1}{2}=z\,d/dz+\tfrac{1}{2}$.

The commutator of the operators $a$ and $a^+$ equals unity ($[a,a^+]=1$),
whence, taking into account that $z=(q-ip)/\sqrt{2}$, i.e., that
$\hat{q}=(a+a^+)/\sqrt{2}$, $\hat{p}=(a-a^+)/i\sqrt{2}$, we find:
$[\hat{q},\hat{p}]=i$,
$\hat{H}=(\hat{q}^2+\hat{p}^2)/2$.
In the Appendix the commutators are obtained with explicit account of
$\hbar$ ($[a,a^+]=\hbar$ and $[\hat{q},\hat{p}]=i\hbar$).  Thereby the
quantum-mechanical apparatus of the harmonic oscillator is reproduced.
The classical equations of motion in the model turn out to be identical to
the quantum ones.

According to~\eqref{eq:4.2} the Poisson bracket equals
\begin{equation}\label{eq:4.11}
\{f,g\} = \omega(\bar{z},z)\left(
  \frac{\partial f}{\partial z}\frac{\partial g}{\partial\bar{z}}
- \frac{\partial f}{\partial\bar{z}}\frac{\partial g}{\partial z}
  \right),\qquad
\omega^* = e^{\beta z\bar{z}}.
\end{equation}

From here we have
$\dot{z}=\{z,H\}=ie^{\beta z\bar{z}}(-\bar{z})$
(i.e., $\dot{z}=ie^{\beta z\bar{z}}(-1)[\bar{z},H]_-=-ie^{\beta z\bar{z}}
\bar{z}$), i.e.,
\begin{equation*}
\dot{z} = -i\omega z,
\end{equation*}
since $\{e^{\beta z\bar{z}},H\}=0$; here $\omega^*=e^{\beta z\bar{z}}$.

Since analytic functions are determined by their values on the boundary of
a region, one can pass to an equivalent description with the help of
complex random variables on the real axis $\Psi_n(q)$.  Introducing the
eigenfunctions of the operator $\hat{H}$ on the $q$-axis:
$\langle q|n\rangle=\Psi_n(q)$, where $\Psi_n(q)$ is the orthonormalized
system of Hermite functions, one can construct the kernel of the unitary
operator $U(z,q)$ connecting entire complex functions $f(z)$ of Fock space
with functions $\psi(q)$ from $L_2$ on $\mathbb{R}_1$:
\begin{equation}\label{eq:4.12}
U(z,q) = \sum_n Z_n(z)\,\Psi_n(q).
\end{equation}

The function $U(z,q)$ possesses the following properties (Section~10.4):
\begin{equation*}
\int_{-\infty}^{\infty} dq\;U(z,q)\,\overline{U(z',q)} = e^{z\bar{z}'},
\end{equation*}
the exponential $e^{z\bar{z}'}$ being the kernel of the identity operator
for the scalar product with measure~\eqref{eq:4.8} (``complex
$\delta$-function'').  Further,
\begin{equation*}
\int d\mu(z,\bar{z})\;U(z,q)\,\overline{U(z,q')} = \delta(q-q'),
\end{equation*}
\begin{equation*}
\int_{-\infty}^{\infty} dq\;U(z,q)\,\Psi_n(q) = Z_n(z),\qquad
\int d\mu(z,\bar{z})\;Z_n(z)\,\overline{U(z,q)} = \Psi_n(q),
\end{equation*}
i.e., the functions $Z_n(z)$ correspond to the eigenfunctions of the
harmonic oscillator on the $q$-axis with energies $E_n$.

Let us note that here the transition from the statistical description
(Gibbs distribution) to the quantum one (Fock space) is not connected with
any ``act of quantization''.  One is dealing only with the transition from
the Gibbs probability distribution to a collection of distributions --- in
complete accordance with standard probability theory (Section~2).  Thus,
the Hilbert space appears as one of the possible schemes of description,
admitted by classical probability theory.  The real ``act of quantization''
will be the recognition of superpositions of random variables as admissible
(physical) states, i.e., the inclusion in the scheme of the superposition
principle of states.

Let us stress that the wave functions $f(z)$ are \emph{dynamical variables}.
This is especially important in connection with Section~3, where wave
functions were identified with functions of excitation of fields --- the
correspondence is complete.  And from the standpoint of Hamiltonian mechanics
the variables $z$, $\bar{z}$ are simpler objects than $q$, $p$ by virtue of
their monochromaticity (in both $q$ and $p$ there are frequencies of both
signs).

\subsection{The classical limit}
\label{sec:4.3}

Let us discuss the meaning of the transition to the classical limit.  Of
course, if one sends $\hbar$ to zero in commutators and other expressions,
one arrives at a well-known result: classical theory.  Now the operators of
coordinate and momentum commute.  The ``wave'' and ``corpuscular'' aspects
of QM then separate.  Waves become only waves, and particles become only
particles.  The transition in the equality $p=\hbar k$ to the limit
$\hbar\to 0$ is possible in two different regimes and leads to two different
pictures:

1) $k\to\infty$, $p\ne 0$;

2) $p\to 0$, $k<\infty$.

Obviously, the first limit gives the corpuscular picture, and the second
gives the wave picture.  Analogously interpreted are also the transitions to
``large quantum numbers'' --- always the limit $\hbar\to 0$ is taken.

It is of interest to consider this question within the framework of the
proposed approach.  According to~\eqref{eq:4.4},
$\hbar=kT/\omega$.  At $\omega=\text{const}$, the limit $\hbar\to 0$
implies $T\to 0$, i.e., freezing of the system.  This is a very
satisfactory conclusion: at zero temperature fluctuations die out and the
dynamics becomes deterministic.  From the standpoint of the fundamental
oscillator, the limit $T\to 0$ means the vanishing of the radius of the
sphere representing the phase space: $h=4\pi R^2$, and $R\to 0$ if
$h\to 0$.  Then the chain of oscillators in Section~5 turns into an ordinary
classical system.  The action of external forces on it generates excitations
propagating along the chain.

It is also not without interest to recall the features of radiation of an
absolutely black body.  The spectral density of radiation is given by the
Planck formula
\begin{equation}\label{eq:4.13}
u(\nu,T) = \frac{8\pi h\nu^3}{c^3}\;\frac{1}{e^{h\nu/kT}-1}.
\end{equation}
At high frequencies, as it should be, we obtain Wien's law
\begin{equation*}
u(\nu,T) = \frac{8\pi h\nu^3}{c^3}\;e^{-h\nu/kT},
\end{equation*}
i.e., the classical corpuscular limit.  At low temperatures the same limit
is obtained.  Conversely, at small frequencies we obtain the Rayleigh--Jeans
formula
\begin{equation*}
u(\nu,T) = \frac{8\pi\nu^2}{c^3}\;kT,
\end{equation*}
i.e., the result of the classical wave theory.  It is also obtained in the
limit of high temperatures.  In this sense the model considered above is in
complete agreement with well-known facts.  One should, however, keep in mind
that the meaning of the parameter $T$ in formulas~\eqref{eq:4.3},
\eqref{eq:4.4} and~\eqref{eq:4.13} is different: in~\eqref{eq:4.3},
\eqref{eq:4.4} $T$ is the temperature of the thermostat in which the
oscillator is placed (it is precisely the thermostat that introduces the
element of randomness into the mechanics of the microworld), while
in~\eqref{eq:4.13} $T$ is the temperature of the absolutely black body.

We conclude: Planck's constant $h$ and the quantum description are easily
modeled in classical mechanics.  But here there are as yet no particles
(quanta), because there is no field.

%======================================================================
\section{Quantum Field Theory}
\label{sec:QFT}
%======================================================================

\subsection{Chain of oscillators}
\label{sec:5.1}

And so, if the PS of a particle is a sphere, then such a system can be
regarded either as a classical Hamiltonian theory of a free particle, or as
a classical oscillator in a thermostat characterized by a Gibbs distribution,
or as a quantum oscillator.  The choice is determined by the formulation of
the problem (physical conditions).  This may be the problem of the motion of
a classical particle (compact PS, Hamiltonian mechanics), or the problem of
describing an oscillator in a thermostat (statistical mechanics).  In the
case of a single oscillator in a thermostat there is no necessity for
passing to the quantum description, and so a model is needed in which there
would be no alternative and probability amplitudes would appear
automatically.  Since, as is evident from Section~3, the problem of the
quantum description of particles reduces to the problem of the quantum
description of fields, it is necessary to construct a model of a quantized
field, i.e., to consider a system with an infinite number of degrees of
freedom.  It turns out that probability amplitudes will be called for upon
passing to an \emph{ordered} system of oscillators, namely, to a chain of
interacting oscillators in a thermostat.  The ordering is precisely that
indispensable circumstance (external condition) which predetermines the
choice of the last of the three listed possibilities.

The equation of motion for small perturbations of such a chain,
characterized by the Lagrangian
\begin{equation}\label{eq:5.1}
L = \sum_n \tfrac{1}{2}\bigl(\dot{q}_n^2 - m^2 q_n^2
  - \gamma(q_{n+1}-q_n)^2\bigr),
\end{equation}
(the constant $\gamma>0$) in the continuous limit coincides with the
Klein--Fock--Gordon equation
\begin{equation*}
\ddot{\varphi} - \varphi'' + m^2\varphi = 0,
\end{equation*}
corresponding to the Lagrangian
\begin{equation}\label{eq:5.2}
L = \tfrac{1}{2}\int dx\,
\bigl(\dot{\varphi}^2 - \varphi'^2 - m^2\varphi^2\bigr).
\end{equation}
(Here $\varphi(x,t)$ is defined by the limiting value of $q_n$ when the
distances between neighboring oscillators $a\to 0$ and $n\to\infty$ so that
$an\to x$, $\gamma a^2\to 1$ and $q_n/a\to\varphi(x,t)$;
$\dot{\varphi}=\partial\varphi/\partial t$,
$\varphi'=\partial\varphi/\partial x$.)  The role of complex amplitudes is
played by solutions of the KFG equation with positive energy
$\varphi^+\sim\exp(-iEt)$, $E>0$, and the probability density is given by
$\rho(x)\sim i(\varphi^{+*}\dot{\varphi}^+-\dot{\varphi}^{+*}\varphi^+)$
(Section~2.2).  In the non-relativistic limit $m\to\infty$ the KFG equation
passes into the one-dimensional Schr\"odinger equation for a free particle:
$\varphi^+=\exp(-imt)\psi$, $i\dot{\psi}=-\psi''/2m$, and
$\rho(x)\sim\psi^*\psi$.

For what follows the following circumstances are essential.  Upon placing the
chain in a thermostat, the ordering of the oscillators plays a fundamental
role.  An arbitrary unordered system of oscillators in a thermostat is
described within the framework of statistical physics.  The necessity of
using a new apparatus is revealed upon passing to a chain of interacting
oscillators.  In this case the information about the excitation of the
system is contained only in one collection of variables: $q_n(t)$ from
$q_n(t)$ and $p_n(t)$, or $\varphi^+(x,t)$ from $\varphi^+(x,t)$ and
$\pi^+(x,t)$; for a single oscillator this is $z$ (or $f(z)$) instead of
$z$ and $\bar{z}$.

Then the scalar product~\eqref{eq:4.9} acquires physical meaning.  The
information about the system is contained in the function $f(z)$ (or,
correspondingly, in $q_n(t)$, $\varphi^+(x,t)$), and for the computation of
probabilities both $z$ and $\bar{z}$ (correspondingly, $q_n(t)$, $p_n(t)$,
$\varphi^+(x,t)$, $\pi^+(x,t)$) are needed, i.e., the functions $f(z)$ and
$\overline{f(z)}$.

From~\eqref{eq:4.9} it follows (taking the oscillator) that in the space of
functions $f(z)$ one can define operators (see Section~10.4):
$a^+ f(z)=zf(z)$, $af(z)=\hbar\,df(z)/dz$ (i.e., $z\to a$,
$\bar{z}\to a^+$), $[a,a^+]=\hbar$.  Since the canonical variables $z$,
$\bar{z}$ become operators, all physical quantities --- i.e., $q_n(t)$,
$p_n(t)$, $\varphi$, $\pi$, and the Hamiltonian $H$ --- also become
operators.  To avoid misunderstanding, let us stress that in~\eqref{eq:4.9}
$z$, $\bar{z}$, $f(z)$, $g(z)$ are ordinary complex variables --- operator
properties appear only upon identifying $f(z)$, $g(z)$ with vectors in
Fock space and upon multiplying them by functions of $z$, $\bar{z}$.

In what follows we shall need the rule by which the Hamiltonian operator
transforms upon changing the standard commutator $[a,a^+]=\hbar$.  Having
in mind the final goal, let us pass to the notation adopted in QFT:
$z\to a$, $\bar{z}\to a^*$, where $a$ is a complex number.  Obviously,
upon the transition $a\to a_\lambda$, $[a_\lambda,a_\lambda^+]=\hbar/\lambda$
(i.e., upon the replacement in~\eqref{eq:4.9}, \eqref{eq:10.2} of
$z\bar{z}/\hbar\to\lambda z\bar{z}/\hbar$), the Hamiltonian takes the form
\begin{equation}\label{eq:5.3}
\hat{H} = \omega\,a_\lambda^+ a_\lambda + \tfrac{1}{2}\omega\hbar,\qquad
a_\lambda = \hbar\lambda^{-1}\,d/da_\lambda^* = a/\sqrt{\lambda},\quad
a_\lambda^+ = a^+/\sqrt{\lambda}.
\end{equation}
The spectra of the Hamiltonians $\hat{H}$ and $\hat{H}'$ coincide.  Such
non-canonical (non-unitary) transformations do not change the content of the
theory and are frequently employed in practical calculations; for example, if
one uses the invariant normalization of states ``$2E$ particles per unit
volume'', then the ``invariant commutation relations''
$[a(\mathbf{k}),a^+(\mathbf{k}')]=(2\pi)^3 2\omega(\mathbf{k})
\delta(\mathbf{k}-\mathbf{k}')$ are convenient.  Let us note that although
the exponent in~\eqref{eq:4.9}, \eqref{eq:10.2} is proportional to the
classical Hamiltonian $\omega a^* a$ ($\beta\omega=1/\hbar$), it cannot be
identified with the operator~\eqref{eq:5.3}.  If in~\eqref{eq:4.9},
\eqref{eq:10.2} one takes the standard exponent $a^* a$, then
in~\eqref{eq:5.3} one should set $\lambda=\hbar$.

Let us show that as a result we obtain a model of a one-dimensional
quantized field.

\subsection{Chain of oscillators in a thermostat}
\label{sec:5.2}

Let us turn to concrete calculations.  Let in the Gibbs distribution
$\exp(-\beta H)$ the Hamiltonian
\begin{equation}\label{eq:5.4}
H = \sum_n \tfrac{1}{2}\bigl(p_n^2 + m^2 q_n^2
  + \gamma(q_{n+1}-q_n)^2\bigr)
\end{equation}
appear.  In normal coordinates $u(k)$, $p(k)$:
\begin{equation}\label{eq:5.5}
q_n = \int_{-\Delta}^{\Delta} dk\;\varphi_n(k)\,u(k),\qquad
\Delta = \pi/a,
\end{equation}
where $\int_\Delta dk\,\varphi_n(k)\,\bar{\varphi}_{n'}(k)=\delta_{nn'}$,
the Hamiltonian is written as
\begin{equation}\label{eq:5.6}
H = \int_{-\Delta}^{\Delta} dk\,
\bigl(p(k)\bar{p}(k)+\tfrac{1}{2}\omega_k^2 u(k)\bar{u}(k)\bigr)
= \tfrac{1}{2}\int_{-\Delta}^{\Delta}
dk\;\omega_k\bigl(a^*(k)a(k)+a(k)a^*(k)\bigr);
\end{equation}
here $\omega_k^2=m^2+4(\gamma/a^2)\sin^2(\pi k/2\Delta)$,
\begin{equation*}
u(k) = \bigl(a^*(k)+a(-k)\bigr)/\sqrt{2\omega_k},\quad
p(k) = i\bigl(a^*(k)-a(-k)\bigr)\sqrt{\omega_k/2}.
\end{equation*}
In new canonical variables $a_k=\sqrt{\lambda_k}\,a(k)$,
$a_k^+=a^+(k)/\sqrt{\lambda_k}$ ($\lambda_k=\omega_k/\omega$,
$\omega=\sqrt{m^2+\gamma/a^2}$), the Hamiltonian~\eqref{eq:5.6} is
rewritten as
\begin{equation}\label{eq:5.7}
\hat{H} = \omega\int_{-\Delta}^{\Delta} dk\;a_k^+ a_k,
\end{equation}
and the measure in phase space takes the form
\begin{equation}\label{eq:5.8}
d\mu = \prod_k \frac{i}{2\pi\hbar}\,da_k^*\wedge da_k\;
\exp\!\left(-\frac{1}{\hbar}\int dk\;a_k^* a_k\right).
\end{equation}
(Here the limits of integration and the overall normalization factor are
omitted, as is customary in the theory of continual integrals.)  For
random variables (functionals $\Phi[a^*]$) the measure~\eqref{eq:5.8}
allows one to define the scalar product
\begin{equation}\label{eq:5.9}
(\Phi_1,\Phi_2)
= \int\prod_k\frac{i}{2\pi\hbar}\,da_k^*\wedge da_k\;
  e^{-\frac{1}{\hbar}\int dk\,a_k^* a_k}\;
  \Phi_1[a^*]\,\overline{\Phi_2[a^*]}.
\end{equation}
According to~\eqref{eq:5.9}, the operators
$\hat{a}_k$, $\hat{a}_k^+$
($\hat{a}_k^+\Phi[a^*]=a_k^*\Phi[a^*]$,
$\hat{a}_k\Phi[a^*]=\hbar\,\delta\Phi[a^*]/\delta a_k^*$)
satisfy the commutation relations
$[\hat{a}_k,\hat{a}_{k'}^+]=\hbar\,\delta(k-k')$, and the Hamiltonian
operator is given by the integral
\begin{equation}\label{eq:5.10}
\hat{H} = \tfrac{1}{2}\omega\int_{-\Delta}^{\Delta}
dk\;\omega_k\bigl(\hat{a}_k^+\hat{a}_k+\hat{a}_k\hat{a}_k^+\bigr).
\end{equation}

Of course, here too one can still consider only problems of statistical
physics, i.e., study only functionals $F[a,a^*]$.

Formula~\eqref{eq:5.9} is nothing other than the scalar product in the Fock
space of a one-dimensional quantum field theory, if in it one takes the
limit $a\to 0$, $\Delta\to\infty$.  Each oscillator of the chain is
quantized (see Section~4), so it is not surprising that such a chain models
a one-dimensional quantum field theory.  Let us once more stress that the
appearance of the description with the help of probability amplitudes is
connected with the fact that one studies the propagation of perturbations
along the chain, i.e., in the theory only $q_n$ figure --- only the
coordinates of neighbors ``interact''; the momenta of neighbors ``do not
interact''.

To summarize: one-dimensional quantum theory of a scalar field is modeled by
a chain of classical oscillators in a thermostat, provided that the Gibbs
measure is identified with the volume measure of phase space.

The model is easily generalized to the multi-dimensional case.  For example,
if the oscillators are placed at the nodes of a $D$-dimensional lattice,
$D>1$, then in the limit $a\to 0$ one obtains a quantum theory of a scalar
field in a space of dimension~$D$.  Very significantly, this structure models
\emph{physical space}: space is formed by a mechanical system characterized
by a Lagrangian.  The ground state of the system is the vacuum (``empty
space''), single-particle excitations of the field are particles described
by quantum mechanics.  For a more realistic model one should pass to
superstrings (in order to include fermions), place them in a
multi-dimensional space, and construct from them a multi-dimensional
network~\cite{Prokhorov2002,ProkhorovISHEPP}.

Thus, for modeling quantum field theory one must go beyond the framework of
physical space; in fact, both quantum mechanics and space are simultaneously
modeled.  Characteristic features of the model: 1)~Planckian distances
($a\sim l_P$), 2)~discreteness, 3)~classicality, 4)~the possibility of
dissipative processes.

%======================================================================
\section{Macroscopic Bodies and the Measurement Problem}
\label{sec:macro}
%======================================================================

\subsection{The measurement problem}
\label{sec:6.1}

Questions connected with measurement occupy a special place in quantum
physics.  Their peculiarity is that they concern an entirely new aspect of
the theory --- the act of measurement, the possibility of whose description
is not obvious, while the importance for quantum physics is enormous.  It
is necessary to know exactly what information an experiment can give.  The
problem arose and acquired sharpness already in the period of the creation
of QM\@.  It was clear that the new mechanics differed radically from
classical mechanics: unusual objects, possessing both wave and corpuscular
properties, were described with the help of an unusual mathematical
apparatus --- probability amplitudes, vectors from a Hilbert space.  Between
the new and old mechanics there yawned an abyss; they were connected only
by a thin thread --- the correspondence principle: in limiting cases (small
$\hbar$, large quantum numbers) quantum theory passed into the classical
one.  But in the act of measurement there participate representatives of
both mechanics: a micro-object, subject to QM, and a classical apparatus.

Moreover, unlike a classical particle, a micro-object cannot simultaneously
possess definite values of, say, coordinate and momentum, and therefore the
physicist at his discretion decides what to measure.  Depending on his
choice, this or that equipment is used.  It creates the impression that
it is precisely the experimenter who effects the transition from the
possible to the actual.

Which mechanics should be used to describe the act of measurement?
\emph{A priori} there is an alternative: (a) the apparatus does not obey the
laws of QM; (b) the apparatus is described by a wave function.  At the dawn
of quantum physics only the first possibility was admitted (the classical
character of apparatus was ``obvious''), which generated numerous
difficulties.  The tacit recognition of the impossibility of describing the
act of measurement either in quantum (due to the presence of the apparatus)
or in classical (due to the presence of the micro-object) theories led to
hypotheses about ``uncontrollable'' interaction of the object with the
apparatus, abrupt (``non-causal''~\cite{Messiah}) change of the wave
function in the process of measurement, etc.

Another source of difficulties were the probability amplitudes.  The
outcome of a measurement must be a distribution of probabilities.  But the
micro-object is described by a wave function, whose evolution is determined
by the Schr\"odinger equation.  The question arises: how, in the process
of measurement, is the transition from probability amplitudes to
probabilities effected?  The answer was the recognition of the outstanding
role of the experimenter.  It was proposed that it is precisely he, having
``looked at the apparatus'', who transforms the ``potential
possibilities''\footnote{The term was introduced by V.\,A.~Fock~\cite{Fock}.
Here ``potential'' means existing in potentia, hidden.}, inherent in the
probability amplitudes, into probabilities proper.  In fact, it was proposed
that precisely his intervention in the experiment in the capacity of
observer effects the transition $\psi\to|\psi|^2$.  Hence the ideas about
the important role of ``consciousness'' in quantum
physics~\cite{Wigner40}.

The difficulty is that here one needs to know the answers to no fewer than
four fundamental questions: about the nature of the observed objects, about
the specifics of the mathematical apparatus of QM (the first two questions
from Group~II), about the specifics of the quantum description of apparatus
(macroscopic bodies), and finally, about the possibility of describing the
act of measurement (questions~1 and~3 from Group~III).

Thus, three problems were superimposed at once: the complexity of the
observed object (from the standpoint of classical physics), the complexity of
the apparatus (from the standpoint of QM), and the incompleteness of the
mathematical apparatus of QM as a branch of probability theory (the place
of probability amplitudes in probability theory was not indicated).  The
complexity of the apparatus generates the question of the possibility of
describing the process of measurement.

Meanwhile, the fact that apparatus, like any other macroscopic bodies, must
obey the laws of QM and their states must be described by wave functions is
sufficiently obvious.  This automatically follows from the fact that any
macroscopic body is a many-particle excitation of fields and is by
definition described by a state vector.  But if a macroscopic body is
described by a wave function, then the act of measurement must proceed in
accordance with the laws of QM: the system object $+$ apparatus is
described by a wave function changing in the process of measurement
according to the Schr\"odinger equation.  There are no jumps, no
``uncontrollable interactions'', no ``non-causal reductions''.

At the present time it is recognized that macroscopic bodies and the
process of measurement can be described within the framework of QM\@.
However, two exceedingly difficult problems remain: the mechanism of the
transformation of probability amplitudes into probabilities
($\psi\to|\psi|^2$), and the absence of superposition of macroscopically
distinct states of macroscopic bodies (apparatus).

The first of these is sufficiently serious.  The evolution operator of the
system is linear, and therefore one must explain how $|\psi|^2$ appears in
the process of measurement.  The answer is given by the axioms of Section~2.
Quantum mechanics is an intrinsically probabilistic science, i.e., for each
event probabilities intrinsically exist and can be indicated.  The
peculiarity of QM as a stochastic theory is that the evolution equations are
formulated not for the probabilities themselves but for wave functions
$\psi$.  How the probabilities change with time is determined by the
Schr\"odinger equation.  From the standpoint of probability theory, the
role of experiment in QM is the same as in classical theory.

The second problem is as follows.  The act of measurement reduces to the
process
\begin{equation}\label{eq:6.1}
|o\rangle|A\rangle \to \sum_k c_k\,|k\rangle|A_k\rangle,
\end{equation}
where $|o\rangle$ and $|A\rangle$ are normalized vectors of the initial
states of the object and the measuring device, and $|k\rangle$, $|A_k\rangle$
are vectors of their possible final states.  According to QM, the object and
the apparatus in the final state are described by a certain wave function.
Meanwhile, a measurement can be considered successful only if the apparatus
turns out to be in just one of the possible states $|A_k\rangle$.  In other
words, the measurement process is carried out under the assumption that for
macroscopic bodies (apparatus) superpositions of macroscopically distinct
states are impossible.  Schr\"odinger~\cite{Schrodinger} dramatized the
situation by choosing a warm-blooded animal as the macroscopic body and life
and death as possible outcomes (``superposition of a live and a dead cat'',
see Section~10.2).  Below it will be shown (Section~6.2) that for
macroscopic bodies there exist superselection rules forbidding superpositions
of macroscopically distinct states.  Then the measurement process looks as
follows: the final state always corresponds to only one of the terms in the
sum $\sum_k c_k|k\rangle|A_k\rangle$, say $|k\rangle|A_k\rangle$, and the
relative frequency of appearance of the state $|A_k\rangle$ is given by
$|c_k|^2$; no additional external agents in the form of ``consciousness''
or an ``observer'', thanks to whose presence a ``jump'' $c_k\to|c_k|^2$
would be effected, are required.  Let us discuss these questions in more
detail.

\medskip
\noindent\textbf{The essence of the problem.}

The main difficulty of the measurement problem is, speaking in modern
language, the problem of \emph{decoherence} of macroscopic bodies
(apparatus), i.e., the transformation of pure states into mixed ones.  A
pure state is described by a vector from a Hilbert space; for a mixed state
only the probabilities with which the system can be found in one or another
pure state are indicated.  Mixed states are conveniently described with the
help of the density matrix ($\rho$-matrix):
\begin{equation}\label{eq:6.2}
\rho = \sum_k p_k\,|k\rangle\langle k|,\qquad
\sum_k p_k = 1,
\end{equation}
where $p_k$ are the probabilities of finding the system in the state~$k$.

Two circumstances are of fundamental importance.

\noindent\textbf{A.}\ Macroscopic bodies obey the laws of QM and are
described by state vectors.

\noindent\textbf{B.}\ The act of measurement obeys the laws of QM and is
described by a transition amplitude.

If the initial states of the object and the measuring device are represented
by the vectors $|o\rangle$ and $|A\rangle$, then the measurement process is
depicted by the transition~\eqref{eq:6.1}.  The final state
in~\eqref{eq:6.1} is called \emph{entangled}.  It is easy to verify that
it cannot be represented in the form $|o\rangle'|A\rangle'$, where
$|o\rangle'$, $|A\rangle'$ are certain state vectors of the object and the
apparatus.

The basic principle of the theory of measurement states: the initial and
final states of the apparatus are macroscopically definite.  The apparatus
cannot be in a superposition of macroscopically distinct states, for
example, states with a pointer indicating 0 and 1.  First, such states have
never been observed; second, if such a thing occurred, the apparatus would
not fulfill its function.  Therefore the measurement process associated with
the transition~\eqref{eq:6.1} must include one more indispensable element
--- the process of decoherence, i.e., the process
\begin{equation*}
\sum_{k,k'} c_k c_{k'}^*\,|k\rangle\langle k'|\otimes
|A_k\rangle\langle A_{k'}|
\;\longrightarrow\;
\sum_k |c_k|^2\,|k\rangle\langle k|\otimes|A_k\rangle\langle A_k|.
\end{equation*}

Meanwhile, according to~A, the initial state of the system object $+$
apparatus is represented by the vector $|o\rangle|A\rangle$; according
to~B, the transition~\eqref{eq:6.1} is effected by applying the unitary
evolution operator to it, i.e., the final state is also pure.  But then the
apparatus is not in a macroscopically definite state, and from here it was
concluded that QM does not describe the interaction of object with
apparatus, and even the hypothesis was put forward that the universe
``branches'', with the apparatus in states $|A_k\rangle$ and $|A_l\rangle$
ending up in different universes (``many-worlds
interpretation''~\cite{Everett}; its meaning is that, on the one hand, the
measurement process is described by the Schr\"odinger equation, and on the
other hand it is ``explained'' why the experimenter sees only one of the
possible states of the apparatus).  A detailed calculation of the
process~\eqref{eq:6.1} for real experiments is absent because of its
complexity.

It would seem that the way out is that in practice macroscopic bodies are
never in pure states --- they are always in mixed states.  The environment
always ``measures'' the macroscopic body, causing decoherence~\cite{Zeh}.
But this only shifts the problem.  The system macroscopic body $+$
environment is also described by a wave function, and the argument can be
repeated.

\subsection{Quantum description of macroscopic bodies}
\label{sec:6.2}

A fundamental role is played by the fact that for macroscopic bodies
\emph{superselection rules}~\cite{ProkhorovUFN} exist that forbid
superposition of macroscopically distinct states.

When one says that $\psi_1$ and $\psi_2$ are vectors from the same Hilbert
space, one means that the transition $\psi_1\to\psi_2$ is possible.
Otherwise, both vectors can be taken from the same space only formally.

For macroscopic bodies: a) the transition from one macroscopically definite
state to another requires a macroscopic time (in comparison with
characteristic times of the measuring apparatus); b) a macroscopic number of
degrees of freedom participate in the transition.  Therefore two
macroscopically distinct states of a macroscopic body cannot be connected
by a transition in a finite time.  This means that their superposition is
forbidden --- a \emph{superselection rule} for macroscopic bodies
(see~\cite{ProkhorovUFN} and Section~10.1).

As a consequence, the state of a macroscopic body in which it is found at
the end of measurement is always macroscopically definite, and no
intervention of ``consciousness'' is needed.

Another important consequence: if for macroscopic bodies there exist
superselection rules, then for microscopic objects there are none.  The wave
function of a micro-object is a genuine state vector from a Hilbert space;
superpositions of any states are admissible.

We note that decoherence by the environment~\cite{Zeh} and superselection
rules for macroscopic bodies are two different (though related) mechanisms.
The former is an approximate, dynamical effect; the latter is exact and
follows from the structure of the theory.

The superselection rules for macroscopic bodies resolve the main paradoxes
of the theory of measurement.

%======================================================================
\section{Some Experimental Consequences}
\label{sec:exper}
%======================================================================

\subsection{Splitting of quanta}
\label{sec:7.1}

If a particle is a non-local excitation of a field, then a natural question
arises: can a quantum be ``split'', i.e., registered in two detectors
simultaneously?  In the integer quantum Hall effect~\cite{vonKlitzing} the
quanta are electrons; they are never split.  In the fractional quantum Hall
effect~\cite{Tsui} the situation is less straightforward: the quanta are
excitations with fractional charge $e/3$~\cite{dePicciotto,Reznikov}.  But
here the field whose excitation constitutes a quasiparticle of charge $e/3$
is a many-body field of interacting electrons in a strong magnetic field, not
the electromagnetic field.  The question of whether a photon or electron can
``split'' when passing through a beam splitter has been investigated: the
answer is no --- a single quantum is always detected in only one
detector~\cite{Prokhorov2001,ProkhorovISHEPP,Dirac}.

\subsection{The Hanbury Brown--Twiss experiment}
\label{sec:7.2}

The Hanbury Brown--Twiss (HBT) effect~\cite{HBT} is the observation of
correlations in the intensities of light from two separate detectors
illuminated by the same thermal source.  Classically, this is understood
as intensity correlations of a classical electromagnetic field.  In the
quantum picture, it is interpreted as photon bunching: photons (bosons) tend
to arrive together.

Recently, single-photon sources have made it possible to test the
indivisibility of photons directly.  Experiments~\cite{Santori} with single
photons emitted one at a time show perfect anti-correlation at a beam
splitter, confirming the integrity of the quantum: a single photon goes to
one detector or the other, never to both.

The HBT effect and single-photon experiments together confirm the picture
emerging from QFT: a photon is an indivisible quantum of the electromagnetic
field, yet the correlations in multi-photon light are precisely those
predicted by the quantum theory of radiation.

%======================================================================
\section{Some Discussion Questions}
\label{sec:discussion}
%======================================================================

\subsection{Various interpretations of quantum mechanics}
\label{sec:8.1}

Orthodox QM does not give satisfactory answers to a whole range of
fundamental questions, in connection with which the latter acquire the
status of paradoxes (Einstein--Podolsky--Rosen~\cite{EPR},
Schr\"odinger (superposition of ``a live and a dead cat'')~\cite{Schrodinger}),
or even problems (the measurement problem, the problem of the quantum
description of macroscopic bodies, etc.).  To these questions one cannot
give answers within the framework of canonical non-relativistic quantum
mechanics.  For this, additional considerations and even going beyond its
framework (invoking QFT) are required.  The incompleteness of the
generally accepted formulation of QM stimulated the search for other
approaches, alternative viewpoints.  Let us enumerate and briefly
characterize the most popular ``interpretations''.

\begin{enumerate}
\item Canonical (or Copenhagen) interpretation (CI).
\item Statistical interpretation (SI)~\cite{Ballentine}.
\item Many-worlds interpretation (MWI)~\cite{Everett}.
\item Hydrodynamic interpretation (HI)~\cite{Madelung}.
\end{enumerate}

\medskip
\noindent\textbf{CI.}\ Practically universally accepted.  It is based on the
assertion that the wave function describes an individual system (particle).
Unfortunately, there is no canonical text (something like a quantum Bible)
containing the Copenhagen interpretation.  The only authoritative and
complete sources touching upon fundamental problems and written by
physicists who took part in the creation of QM are the books
\cite{Dirac,Heisenberg,Fock} and the article~\cite{BohrNature}.  But even
they cannot be considered final, since the views of the founding fathers
changed with time.  For example, initially N.~Bohr assumed that the quantum
postulate ``\ldots includes renunciation of causal spatio-temporal
description of atomic processes''~\cite{BohrNature}; that ``\ldots the
theory of relativity preserves causal description; in quantum mechanics we
are forced to renounce this, to renounce because of the uncontrollable
interaction between objects and measuring apparatus''~\cite{BohrBook}.
Later (after the meeting with Fock) he already wrote about ``\ldots broader
frameworks of causality'' in QM~\cite{BohrBook}, about ``\ldots a
fundamental distinction between measurement instruments and the phenomena
under investigation by means of those instruments'' (ibid.).  So far as
linearity is concerned, Weinberg~\cite{Weinberg} showed that in the
presence of nonlinearity the theory leads to unacceptable physical
consequences (superluminal signal transmission).  Alternative
interpretations~\cite{Skorobogatov,SkorobogatovIJTP} of Weinberg's result
are also possible.

\medskip
\noindent\textbf{SI.}\ The viewpoint of the statistical interpretation is
directly opposite: the wave function describes ``statistical properties of
an ensemble of systems''; it ``does not give a complete description of an
individual system''~\cite{Ballentine}.  This implies that QM is incomplete.

\medskip
\noindent\textbf{MWI.}\ Everett's interpretation~\cite{Everett} proceeds
from the indisputable assumption that all bodies in nature obey QM.  The
measurement process
$|o\rangle|A\rangle\to(|m\rangle|A_m\rangle+|n\rangle|A_n\rangle)/\sqrt{2}$
is analyzed.  Since the final state is a superposition, the apparatus is not
in a macroscopically definite state.  This leads to the ``many-worlds''
hypothesis.

The reason for the appearance of MWI is the insufficient development of the
quantum mechanics of macroscopic bodies.

\medskip
\noindent\textbf{HI.}\ Based on the similarity of the equations of QM and
hydrodynamics~\cite{Madelung}.  In the non-relativistic Schr\"odinger
equation one passes to new functions:
$\psi=\sqrt{\rho}\,e^{iS/\hbar}$.  For $\rho$ and $S$ one obtains
equations:
\begin{equation*}
\frac{\partial\rho}{\partial t} = -\nabla\cdot(\rho\,\mathbf{v}),
\qquad \mathbf{v}=\frac{\nabla S}{m},
\end{equation*}
\begin{equation*}
\frac{\partial S}{\partial t}
= -\frac{(\nabla S)^2}{2m} - V - V_q,
\end{equation*}
the first being the continuity equation, and the second resembling the
Hamilton--Jacobi equation with a ``quantum potential'' $V_q$.

\subsection{Determinism and causality}
\label{sec:8.2}

The absence of classical determinism in QM was no less of a shock than
wave--particle duality.  The problem is resolved already within orthodox QM:
the wave function obeys a causal equation, i.e., in QM there is causality,
but no determinism.  The problem is completely removed if one takes into
account that QM can be viewed as a special case of the theory of random
processes in standard probability theory (Section~2).

\subsection{On language}
\label{sec:8.3}

One sometimes says that in describing the microworld one has to abandon the
language of classical physics.  Is this so?

The language, of course, changed: new words were added (wave function,
quantum, probability amplitude, boson, fermion, \ldots).  But at the
\emph{root} level, i.e., if one takes only the key words, the changes are
minimal for the language and revolutionary for physics.  In fact, only two
new word-combinations appeared: \emph{probability amplitude} and
\emph{superposition of probability amplitudes}.  In everything else one
could make do with the old language and old mathematics.  No change in
logic is required.  The difficulties that arose are connected not with the
language but with the physical content of the new theory: the unusual
properties of micro-objects, the absence of models, the incomplete
(axiomatic) formulation of the theory.  ``Language difficulties'' were
generated by the absence of understanding, and not vice versa.

\subsection{Physics and philosophy}
\label{sec:8.4}

The shock was great.  The instinct of self-preservation generated a new
commandment of the schoolboy: first we do not understand, and then we get
used to it.  ``I think I can safely say that nobody understands quantum
mechanics''~\cite{Feynman}.

The situation was unique.  Essentially it was a profound ideological crisis
against the background of a flawlessly working mathematical apparatus.
The appeal to philosophy was not accidental.  Bohr's complementarity
principle was a concrete embodiment of the philosophy of S.~Kierkegaard and
his student H.~H\o ffding.  ``The uncertainty principle itself was not hard
to formulate, but the consequences it entailed shook the physicists to the
core,'' wrote de~Broglie~\cite{deBroglie}.

But philosophy could play only the role of an analgesic, easing the pain
but not curing the disease.  Only a model could serve as an answer to the
challenge.  A philosophical discussion may be useful, but only to the
extent that it stimulates the search for a model.  Without a model, all
the ``interpretations'' of QM remain hanging in the air --- they explain
the incomprehensible through the unknown.  The model proposed in
Sections~4,~5 of this book is an attempt to answer the challenge.

%======================================================================
\section{Conclusion}
\label{sec:conclusion}
%======================================================================

From the above it follows that the commonly accepted formulation of QM does
not give answers to a whole range of questions.  Some questions are of a
programmatic character and serve as sources of paradoxes and new
interpretations.

\begin{enumerate}
\item What is a particle?
\item What is the wave function?
\item Are wave functions that are nonzero only in non-overlapping regions
of space admissible?
\item Is superposition of macroscopically distinct states of macroscopic
bodies possible?
\item What is the probability amplitude from the standpoint of probability
theory?
\end{enumerate}

Each of them conceals the possibility of an alternative interpretation of
QM\@.  The absence of a clear answer to the first question led to the
statistical interpretation; to the second --- to the hydrodynamic
interpretation.  Questions~2 and~3 generated the Einstein--Podolsky--Rosen
paradox~\cite{EPR}.  Question~4 is of fundamental importance for the theory
of measurement; it generated Schr\"odinger's paradox~\cite{Schrodinger}
and the many-worlds interpretation~\cite{Everett}.

\medskip
Among the most reliably established facts of physics is that all known
particles are quanta (single-particle excitations) of corresponding fields,
and the surrounding world is their manifestation.

The key significance of QFT was not always appreciated.  A solid body serves
as a visual model of quantum field theory: there exist both acoustic and
spin waves in it; the corresponding quanta are called phonons and magnons,
and the mathematical apparatus employed is identical to that of quantum field
theory.  Particles are described by ``wave functions'' characterizing the
deviation of the corresponding degrees of freedom from the equilibrium
position.

Drawing the proper conclusions from here was hindered by many reasons,
including the absence of satisfactory answers to the questions formulated in
the Introduction.

\medskip
\emph{The measurement problem} is second in importance.  But it is closely
connected with the first and with the question of what a particle is.  It
was precisely in connection with measurement that the question arose about
the role of ``consciousness'' in quantum theory.  W.~Heisenberg justified
this as follows~\cite{HeisenbergPP}: ``The probability function combines
objective and subjective elements.  It contains statements about
possibilities or better tendencies (`potentia' in Aristotelian philosophy),
and these statements are completely objective; they do not depend on the
observer; and it contains statements about our knowledge of the system,
which, of course, is subjective, since it may be different for different
observers.''  According to Heisenberg~\cite{HeisenbergPP}: ``it is the
observation that effects the transition of the `possible' into the
`actual' \ldots\ this is not connected with the act of registration of
the result by the brain of the observer.''  At the same time, approximately
at the same period, he wrote~\cite{HeisenbergDaedalus}: ``The laws of
nature, which we formulate mathematically in quantum theory, no longer deal
with the particles themselves but with our knowledge of the elementary
particles.''

E.\,P.~Wigner approached the question of the role of consciousness in
physics somewhat differently~\cite{WignerSR}: ``\ldots one cannot
formulate the laws of quantum mechanics in a fully consistent way without
reference to consciousness.  All that quantum mechanics does is to provide
probability connections between subsequent impressions (also called
`apperceptions') of consciousness.''  And further: ``\ldots the very study
of the external world leads to the conclusion that the content of
consciousness is the ultimate reality.''  As a result~\cite{WignerSR}:
``Solipsism may be logically consistent with present quantum mechanics.''
However, Wigner immediately adds~\cite{WignerSR}: ``\ldots logically, it
is possible to deny the existence of the external world --- although it is
not very practical to do so.''  As applied to the act of measurement, this
philosophy leads to the conclusion~\cite{WignerSR}: ``\ldots the
impression received during interaction, also called the \emph{result of
observation}, modifies the wave function of the system.  \ldots The
fixation of the impression in our consciousness --- this is what modifies
the wave function, because it modifies our assessment of the probabilities
for the various impressions which we expect to receive in the future.''
Remarkably, the starting point of Wigner's reasoning was the Cartesian
``Cogito ergo sum'' (I think, therefore I am)~\cite{WignerSR}.  But
thinking does not reduce to perception alone; existence is impossible
without activity.  It is not enough to receive impressions; one must also
``think'', i.e., structurally process impressions, model the external
world, and employ logic.  The mind is given in order to foresee the
future (I.\,P.~Pavlov).

A similar point of view can be encountered even today.  In~\cite{MenskyMeasurements} we read: the measurement problem ``can be
solved (if this is possible at all) only by including such a concept as
consciousness.''  The hypothesis was even considered of a volitional leader
(an ``experimenter--miracle-worker'') who can influence the results of an
experiment.  Recently, one such miracle-worker was dismissed from his
position for fabricating data on element~118~\cite{Schwarzschild}.
Einstein expressed himself on this matter as follows~\cite{WignerEinstein}:
``If a mouse looks at the universe, does the state of the universe change
because of that?''

\medskip
\noindent\emph{Quantum mechanics and space.}

That the concept of quanta must be connected with the concept of space was
the cornerstone of Einstein's ``unified theory'' programme; in the end, he
began to incline toward the inevitability of abandoning continuity.  In
1954 he wrote to M.~Besso~\cite{EinsteinSbornik}: ``But I consider it
quite possible that physics cannot be founded solely on the concept of the
field, i.e., on continuous structures.''  This required a revision of the
concept of space.  A theory of ``quantized space-time'' was
proposed~\cite{Snyder}; a theory with a curved momentum space was
studied~\cite{Kadyshevsky}.  There remained but one step to the idea of a
curved phase space.  Finally, 't~Hooft~\cite{tHooft} pointed to the
necessity of a transition to a discrete space at Planckian distances.  It
was implied that in this way one would succeed in solving both the problem
of the quantum description (despite the classical character of the theory)
and the problem of black holes (since dissipative processes would become
permissible).

\medskip
As P.\,A.\,M.~Dirac wrote in his later years~\cite{DiracFQ}: ``To the
question of what is the main feature of quantum mechanics, I am now inclined
to answer that it is not the non-commutative algebra.  It is the existence
of probability amplitudes underlying all atomic processes.''

The possibility of axiomatizing the apparatus of QM within the framework of
probability theory (Section~2) dispels the last doubts about the
fundamentality of the concept of probability amplitude.

If a quantized string is viewed as a chain of classical oscillators in a
thermostat, then the whole world --- space plus all fields --- is a
3-dimensional structure built from classical oscillators in a thermostat.
Remarkably, not only quantum mechanics for micro-excitations of the
network is obtained, but also a unified theory of all fields.

%======================================================================
\section*{Appendix}
\addcontentsline{toc}{section}{Appendix}
%======================================================================

\subsection*{10.1\quad Superselection rules}
\label{sec:10.1}

Superselection rules (SR) forbid the superposition of certain states.
The classical example is the prohibition of superposition of states with
integer ($\psi_b$) and half-integer ($\psi_f$) spin:
$\psi=\psi_b+\psi_f$~\cite{WWW}.  Indeed, under a rotation of the
coordinate system through the angle $2\pi$ we have:
$\psi_b\to\psi_b$, $\psi_f\to-\psi_f$, and the vector $\psi$ goes over
into a new vector $\psi'=\psi_b-\psi_f$, such that
$\psi'\ne e^{i\alpha}\psi$, i.e., after an identity transformation it
changes in a non-trivial manner, which is absurd.

Another example of SR is given by electric charge~\cite{WWW}.  The
existence of this rule follows from the property of gauge invariance of
quantum electrodynamics~\cite{Prokhorov1990}, according to which all
physical operators $F$ (self-adjoint operators constructed from canonical
variables) must be gauge invariants.  The Gauss law
$G(x)=\nabla\cdot\mathbf{E}(x)-j_0(x)=0$ is a first-class constraint.
Denoting $G_\omega=\int d^3x\,\omega(x)\,G(x)$, where $\omega(x)$ is
an arbitrary function, the operator $G_\omega$ vanishes on all vectors of
the physical Hilbert space $\mathcal{H}_{\rm ph}$ and commutes (weakly,
i.e., on all vectors of $\mathcal{H}_{\rm ph}$) with all physical
operators.  The operator $U_\omega=\exp(iG_\omega)$ effects gauge
transformations.  If one includes global gauge transformations in the class
of local ones (i.e., if in $G_\omega$ one lets $\omega(x)\to\text{const}$),
then the superselection rule for electric charge follows immediately.
A vector of a physical state $\psi_{e_i}\in\mathcal{H}_{\rm ph}$ with
charge $e_i$ transforms as $\psi_{e_i}\to\exp(ie_i\omega)\psi_{e_i}$.
The superposition $\psi=\psi_{e_1}+\psi_{e_2}$ under the
transformation goes over into
$\exp(ie_1\omega)\psi_{e_1}+\exp(ie_2\omega)\psi_{e_2}$, i.e., changes
non-trivially, whence the superselection rule for electric charge
immediately follows; moreover, it also follows that the total electric
charge of the universe equals zero~\cite{Prokhorov1990,Prokhorov1994}.

The hallmark of the appearance of SR is the existence in the theory of a
physical operator~$S$ commuting with all other physical operators.  Then
states $\psi_\Sigma=\psi_{s_1}+\psi_{s_2}$, $S\psi_{s_i}=s_i\psi_{s_i}$,
$i=1,2$, $s_1\ne s_2$, are devoid of meaning (unrealizable).  Indeed, if
states of the type $\psi_\Sigma$ existed, then there should exist an
operator~$R$ not commuting with~$S$, whose eigenvectors they would be.
There should also exist an operator~$F$ transforming $\psi_s$ into
$\psi_\Sigma$: $\psi_\Sigma=F\psi_s$.  But then $[F,S]\ne 0$.  There are,
by hypothesis, no such operators; the physical possibility of a real
transformation of the state $\psi_s$ into the superposition $\psi_\Sigma$ is
absent.  This constitutes the content of the superselection rules.

\medskip
\noindent\textbf{ASSERTION.}\ Superpositions of vectors belonging to
unitarily inequivalent Hilbert spaces $\mathcal{H}_1$, $\mathcal{H}_2$
(orthogonal spaces with cyclic ground states) are physically unrealizable.

\emph{Proof.}\ No operator $F$ from the space $\mathcal{H}_1$ can transfer
it to the space $\mathcal{H}_2$ (and vice versa), since this cannot be
done for the vacuum: $F|0\rangle_1\perp|0\rangle_2$, because
${}_{2}\langle 0|F|0\rangle_1={}_{2}\langle 0|F\rangle_1=0$
(since $|F\rangle_1\in\mathcal{H}_1$).  It is not hard to exhibit a
non-trivial operator commuting with all physical operators of both
subsystems, which gives rise to superselection rules.  This is a special
case of the assertion in Section~6.2.

As Einstein wrote~\cite{EinsteinSbornik}: ``\ldots superposition of real
states is nonsense, which is directly seen in the case of `macroscopic
bodies'.''

\subsection*{10.2\quad Paradoxes of quantum mechanics}
\label{sec:10.2}

\noindent\textbf{The Einstein--Podolsky--Rosen paradox}~\cite{EPR}.
Suppose a two-particle system is described by
$\Psi(x_1,x_2)$.  Let $A$, $B$ be two non-commuting self-adjoint operators
for particle~1, with eigenfunctions $u_n$, $v_n$ respectively.  Then
$\Psi=\sum_n\psi_n(x_2)\,u_n(x_1)=\sum_n\phi_n(x_2)\,v_n(x_1)$.
By measuring $A$ (resp.\ $B$) on particle~1, one apparently determines the
state of particle~2 to be $\psi_n$ (resp.\ $\phi_n$) without any
interaction.  This seemed to imply either superluminal influence or
incompleteness of QM.

Bell~\cite{Bell1,Bell2} reduced the problem to the question of the
fulfillment of his inequalities.  Experiment~\cite{Aspect} convincingly
confirmed QM.

\medskip
\noindent\textbf{Schr\"odinger's cat}~\cite{Schrodinger}.
A living cat is placed in a sealed chamber with a device that may kill it
based on the decay of a radioactive atom.  The $\psi$-function of the system
represents a superposition of ``alive'' and ``dead'' states.  The paradox
underscores the fragility of the orthodox QM description of measurement and
of macroscopic bodies.  The resolution lies in the superselection rule:
death involves a macroscopic number of cells, so the alive and dead states
are macroscopically distinct, and their superposition is forbidden.

\subsection*{10.3\quad Quantum theory and Planck's constant $h$}
\label{sec:10.3}

There exist classical theories containing a constant of the dimension of
action (theories with compact PS, see Section~4), and there exist quantum
theories \emph{without} Planck's constant $h$.

The following elementary model illustrates this.  Let space consist of two
points (1,2), time be discrete $t_n=n$.  Three types of theories are
possible:

\begin{enumerate}
\item \textbf{Classical.}  $x_i(t_n)=0$ or $1$.
\item \textbf{Stochastic.}  Non-negative $p_i(t_n)$,
$p_1(t_n)+p_2(t_n)=1$.  A Markov process:
$p_i(t_{n+1})=\sum_j w_{ij}(t_n)\,p_j(t_n)$.
\item \textbf{Quantum.}  Two complex numbers $\psi_i(t_0)$ with
$|\psi_1|^2+|\psi_2|^2=1$.  Evolution via unitary $2\times 2$ matrices:
$\psi_i(t_{n+1})=\sum_j S_{ij}(t_n)\,\psi_j(t_n)$.
\end{enumerate}

This model contains no $h$ yet clearly demonstrates a quantum theory ---
description via probability amplitudes not reducible to a classical
(Markov) process due to the presence of interference terms.

\subsection*{10.4\quad Proofs of formulas of Section~4}
\label{sec:10.4}

The orthonormality of $Z_n(z)=z^n/\sqrt{n!}$ is proved by integration by
parts in the scalar product (for $m\ge n$):
\begin{equation*}
(z^n,z^m) = \int\frac{i\,dz\wedge d\bar{z}}{2\pi}\;e^{-z\bar{z}}\;
z^n\bar{z}^m = n!\,\delta_{nm}.
\end{equation*}

The explicit account of $\hbar$ is produced by passing to the Hamiltonian
$\tilde{H}=\omega z\bar{z}$, $z=(q+ip)/\sqrt{2}$, so that
\begin{equation}\label{eq:10.2}
d\mu \propto dz\wedge d\bar{z}\;e^{-\beta\omega z\bar{z}},\qquad
\beta\omega = 1/\hbar.
\end{equation}
Then $a^+\leftrightarrow z$, $a=\hbar\,d/dz$,
$[a,a^+]=\hbar$, $[\hat{q},\hat{p}]=i\hbar$.

%======================================================================

\end{document}